\documentclass{nature-pre}
\usepackage{times}

\usepackage{lineno}

\usepackage[english]{babel}
\usepackage{lipsum}  
\usepackage{amsmath}
\usepackage{graphicx}
\usepackage{svg}
\usepackage{tabularx,multirow}
\usepackage{longtable,threeparttablex}
\graphicspath{{figs/}}
\usepackage{hyperref}
\hypersetup{
 bookmarks=true,		
 unicode=false,			
 pdftoolbar=true,		
 pdffitwindow=false,		
 pdfstartview={FitH},		
 pdfcreator={pdflatex},		
 pdfnewwindow=true,		
 colorlinks=true,		
 linktoc=page,			
 linkcolor=blue,		
 citecolor=blue,		
 filecolor=blue,		
 urlcolor=blue			
}
\usepackage{soul}
\usepackage[normalem]{ulem}
\usepackage{physics}
\usepackage{array}
\usepackage{enumerate}

\usepackage{comment}
\usepackage{ulem}

\begin{document}

\title{Nuclear Spin-Mediated Relaxation Mechanisms of the V$_\text{B}^-$ Center in hBN}


\author{Chanaprom Cholsuk$^{1,2}$, Tobias Vogl$^{1,2,3,*}$, Viktor Ivády$^{4,5,*}$}

\maketitle

\begin{affiliations}
\item {Department of Computer Engineering, TUM School of Computation, Information and Technology, Technical University of Munich, 80333 Munich, Germany}
\item {Munich Center for Quantum Science and Technology (MCQST), 80799 Munich, Germany}
\item {Abbe Center of Photonics, Institute of Applied Physics, Friedrich Schiller University Jena, 07745 Jena, Germany}
\item {Department of Physics of Complex Systems, Eötvös Loránd University, Egyetem tér 1-3, H-1053 Budapest, Hungary}
\item {MTA–ELTE Lendület "Momentum" NewQubit Research Group, Pázmány Péter, Sétány 1/A, 1117 Budapest, Hungary}
\end{affiliations}

\noindent email: {tobias.vogl@tum.de, ivady.viktor@ttk.elte.hu}

\date{\today}

\vspace{1cm}

\begin{abstract}
The negatively charged boron vacancy (V$_\text{B}^{-}$) defect in hexagonal boron nitride has recently emerged as a promising spin qubit for sensing due to its high-temperature spin control and versatile integration into van der Waals structures. While extensive experiments have explored their coherence properties, much less is known about the spin relaxation time ($\boldsymbol{T_1}$) and its control parameter dependence. In this work, we develop a parameter-free spin dynamics model based on the cluster expansion technique to investigate $\boldsymbol{T_1}$ relaxation mechanisms at low temperature. Our results reveal that the V$_\text{B}^{-}$ center constitutes a strongly coupled electron spin-nuclear spin core, which necessitates the inclusion of the coherent dynamics and derived memory effects of the three nearest-neighbor nitrogen nuclear spins. Using this framework, this work closely reproduces the experimentally observed $\boldsymbol{T_1}$ time at $\boldsymbol{B} = 90$~G and further predicts the $\boldsymbol{T_1}$ dependence on external magnetic field in the $ 0 \leq \boldsymbol{B} \leq 2000$~G interval, when the spin relaxation is predominantly driven by electron–nuclear and nuclear–nuclear flip–flop processes mediated by hyperfine and dipolar interactions. This study establishes a reliable and scalable approach for describing $\boldsymbol{T_1}$ relaxation in V$_\text{B}^{-}$ centers and offers microscopic insights to support future developments in nuclear-spin-based quantum technologies.
\end{abstract}


\section{Introduction}
In solid-state quantum systems, spin qubits have emerged as a crucial resource for advancing quantum technologies, including quantum sensing\cite{10.1038/s41563-024-01887-z,10.1038/s41467-021-24725-1}, quantum memory\cite{10.1103/PhysRevX.6.021040,10.1038/nphys2026,10.1126/science.1220513,10.1063/5.0188597,10.1002/adom.202302760}, quantum computing\cite{10.1063/5.0007444}, and quantum repeaters\cite{10.1038/s41534-022-00637-w}. Among various defect-based platforms, the nitrogen-vacancy (NV) center in diamond has demonstrated remarkable potential, owing to its long spin coherence times and the ability to initialize, coherently manipulate, and optically read out individual electron spins at room temperature\cite{10.1016/j.physrep.2013.02.001,10.1126/science.1139831}. Motivated by these achievements, considerable effort has recently focused on identifying and characterizing other optically addressable spin defects in wide-bandgap materials\cite{10.1038/s41563-024-01887-z,10.1021/acsnano.3c08940,10.1038/s41467-024-51129-8,10.1021/acs.jpclett.3c01475}. In particular, point defects in hexagonal boron nitride have attracted growing interest due to their intrinsically high photon out-coupling efficiency\cite{10.1038/nnano.2015.242}, excellent compatibility with photonic components\cite{10.1088/1361-6463/aa7839,10.1002/adom.202002218}, long-term photostability\cite{10.1021/acsphotonics.8b00127,10.1016/j.physe.2020.114251}, bright and spectrally pure single-photon emission\cite{10.1021/acsnano.3c08940}, as well as various defect choices\cite{10.3390/nano12142427,10.1021/acs.jpcc.4c03404}.

Among these, the negatively charged boron vacancy defect has emerged as a candidate for enabling coherent control of an electron spin and nearby nuclear spins in hBN. The V$_\text{B}^{-}$ center consists of a missing boron atom coordinated by three nitrogen atoms in a D$_{3h}$ symmetry, forming an $S=1$ electronic spin system with a zero-field splitting (ZFS) of approximately 3.5 GHz~\cite{10.1038/s41467-023-44494-3,10.1038/s41563-020-0619-6,10.1038/s41524-020-0305-x}. Recent experimental studies have characterized its spin and optical properties~\cite{10.1038/s41467-023-44494-3,10.1038/s41563-020-0619-6,10.1038/s41524-020-0305-x,10.1126/sciadv.abf3630,10.1021/acs.nanolett.1c02495,10.1038/s41563-022-01329-8}. At room temperature, longitudinal relaxation times have been reported in the range of 15–18~$\mu$s, largely independent of applied magnetic field and host isotope composition~\cite{10.1038/s41467-023-44494-3,10.1126/sciadv.abf3630,10.1021/acs.nanolett.1c02495}. At cryogenic temperatures, $T_1$ extends to 1 ms up to 12.5 ms, depending on the isotope concentration and the magnetic field strength\cite{10.1038/s41467-023-44494-3,10.1126/sciadv.abf3630}. Recent experiments have demonstrated that the V$_\text{B}^{-}$ center remains stable in ultrathin, few-layer hBN flakes, exhibiting robust optically detected magnetic resonance signals and spin polarization, while spin-lattice relaxation appears to depend on the dimensionality of the sample.\cite{durand_optically_2023} However, the comprehensive analysis about dependence of $T_1$ on the external magnetic field remains largely unexplored.

In terms of spin coherence, spin echo times ($T_2$) are strongly influenced by the nuclear spin environment, yielding values of 82.1 ns for natural isotopic abundance hBN\cite{10.1038/s41467-022-33399-2}, 46 ns for h$^{11}$BN, and 62 ns for h$^{10}$BN\cite{10.1038/s41467-022-31743-0}. While most early investigations focused on $^{14}$N-containing systems, recent work has identified hB$^{15}$N as a particularly favorable isotopic configuration, offering improved $T_1$ and $T_2$ times due to reduced hyperfine interactions\cite{10.1038/s41467-023-44494-3,10.1002/adfm.202511300,10.1038/s41699-022-00336-2}.

Despite significant experimental and theoretical progress in understanding decoherence in V$_\text{B}^{-}$ center\cite{10.1038/s41467-023-44494-3,10.1038/s41467-022-31743-0,rizzato_extending_2023,ramsay_coherence_2023,10.1002/adfm.202511300,10.1038/s41699-022-00336-2}, comprehensive theoretical modeling of their spin relaxation dynamics remains limited. In particular, while several computational studies have focused on investigating $T_2$ time in these systems\cite{10.1002/adfm.202511300,10.1038/s41699-022-00336-2,10.1103/PhysRevB.106.104108}, systematic modeling of the $T_1$ time has been comparatively scarce. A thorough understanding of the mechanisms governing $T_1$ is essential, as it ultimately sets an upper bound for the achievable spin coherence time through the fundamental relation $T_2 \leq 2T_1$\cite{10.1088/0034-4885/80/1/016001}. Typically, the $T_1$ relaxation time comprises multiple contributions arising from distinct microscopic processes, including spin–spin and spin–lattice interactions\cite{10.1126/sciadv.abf3630,10.1103/PhysRevB.97.094304,10.1038/s41524-021-00673-8}.
At room temperature, spin relaxation is predominantly governed by spin-lattice interactions\cite{10.1038/s41467-023-44494-3,10.1103/PhysRevLett.108.197601}. This can be theoretically captured by \textit{ab initio} spin dynamics simulations\cite{10.1038/s41524-023-01082-9,10.1103/PhysRevB.98.214442}. However, at cryogenic temperatures, where phonon populations are suppressed, spin relaxation is largely dictated by spin flip-flop processes through hyperfine and dipolar interactions with the surrounding nuclear spin bath. Capturing dipolar spin relaxation is particularly challenging, as the approximations used for computing decoherence in cluster correlation expansion methods are not applicable, and many-spin entanglements may need to be taken into account, especially in the dense nuclear spin environment in hBN\cite{10.1103/PhysRevB.101.155203}.  To date, no comprehensive model has been constructed to fully capture these low-temperature spin relaxation mechanisms in V$_\text{B}^{-}$ centers. 

In this work, we therefore develop a spin dynamics model to investigate the $T_1$ relaxation mechanism of V$_\text{B}^{-}$ centers in bulk hBN at low temperature and moderate magnetic fields. To accurately account for the interplay of hyperfine interactions, dipolar couplings, and electron-nuclear entanglement effects, we employ a cluster expansion technique combined with an extended Lindbladian formalism\cite{10.1103/PhysRevB.101.155203,xu_ab_2024}. This approach enables us to approximate the time evolution of the diagonal elements of the reduced density matrix of the central spin system without simplifying the underlying spin interactions, while remaining computationally tractable. Using this framework, we first carry out various models and then identify the most effective model capable of accurately capturing spin decoherence. We show that the electron spin-three first-neighbor $^{15}$N nuclear spin coupled system must be treated as a single coherent unit, an extended core, in order to reproduce the experimentally observed decay characteristics. We then apply this optimized model to elucidate the relaxation mechanisms responsible for the experimentally observed $T_1$ at $B = 90$~G\cite{10.1038/s41467-023-44494-3}. Finally, we extend our analysis to predict the $T_1$ dependence on the external magnetic field.  These results offer valuable microscopic insight into the spin relaxation dynamics of V$_\text{B}^{-}$ centers and establish a theoretical foundation for future developments in coherence control, defect engineering, and device optimization for quantum information applications.

\section{Results}
The analysis of $T_1$ relaxation time is structured into three different subsections. First, we investigate the variations in the $T_1$ values derived from four computational models, which include memory effects of the bath on different levels. The relevance of this part is to gain deeper insight into the decay mechanism of the V$_\text{B}^{-}$ center and to identify a suitable, feasible model for computing $T_1$. Next, we assess  predictions of the optimal model against the experimental measurements. Finally, we examine the dependence of $T_1$ time on the external $B$ strength.

\begin{figure*}
    \centering
    \includegraphics[width=1\linewidth]{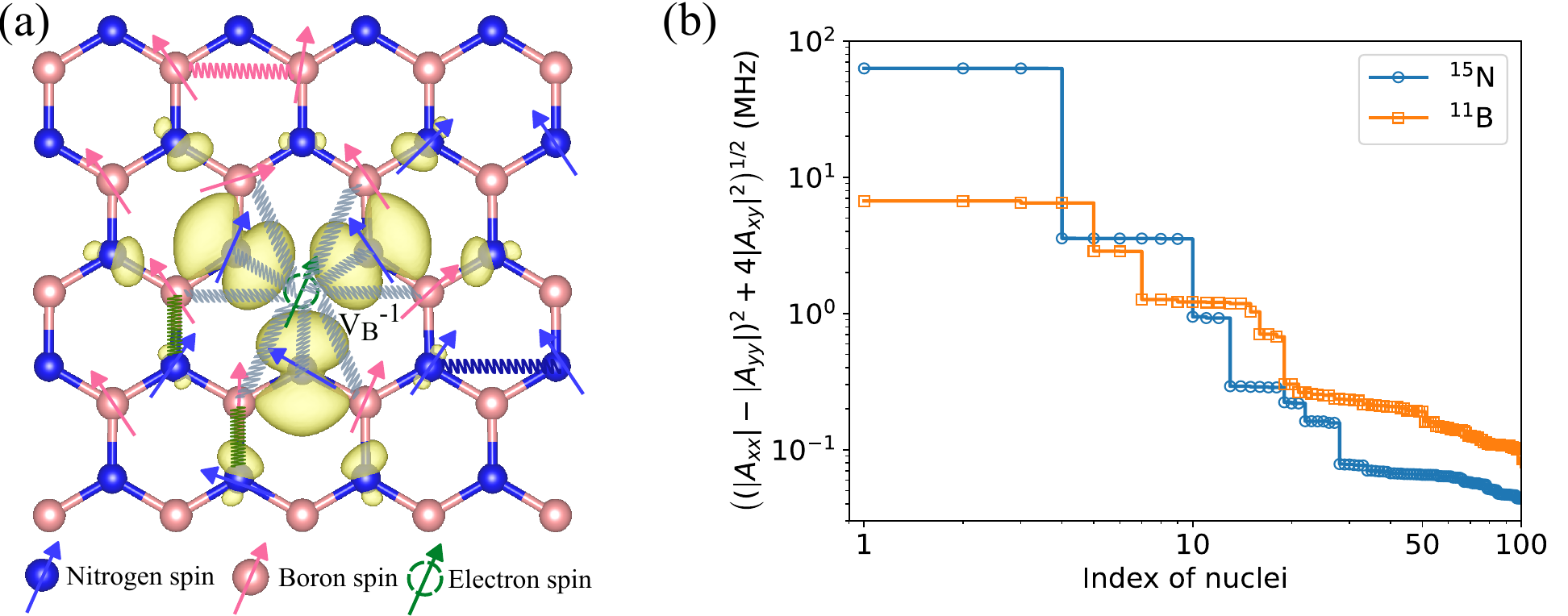}
    \caption{\textbf{Possible interactions of a V$_\text{B}^{-}$ defect in hBN.} (a) Spin density of the V$_B^{-}$ defect, highlighting various interactions. The hyperfine interactions (Eq.~\eqref{eq:H_SB}) are represented by gray wavy lines, while the dipole-dipole interactions between nuclear spins, including nitrogen-nitrogen, boron-boron, and nitrogen-boron interactions, are depicted by blue, pink, and green wavy lines, respectively (corresponding to the second term in Eq.~\eqref{eq:H_B}). The interactions of both electron and nuclear spins with an external magnetic field are present but not explicitly visualized in this figure.  The spin density illustrates the localized wavefunction, which correlates with the strength of the hyperfine interaction. (b) Rotational invariant perpendicular hyperfine components in decreasing order plotted in a log-log scale. The steps indicate different neighboring shells. The first three nearest-neighbor nitrogen nuclei exhibit an extraordinarily strong hyperfine interaction, indicating a strongly induced spin flip-flop process.}
    \label{fig:vb-1_interaction}
\end{figure*}

To investigate the spin relaxation mechanism of the V$_\text{B}^{-}$ defect in h$^{11}$B$^{15}$N bulk under low temperature, we focus on spin relaxation induced by interactions between the electron of the  spin of the V$_\text{B}^{-}$ and the nuclear spin bath, as depicted in Fig~\ref{fig:vb-1_interaction}(a). These interactions can be described by the following spin Hamiltonian
\begin{equation}
    H = H_{S} + H_{B} + H_{SB}, \label{eq:total_H}
\end{equation}
where \(H_S\) accounts for the Hamiltonian for the central spin; $H_{B}$ represents the Hamiltonian for the bath spins in the environment; and \(H_{SB}\) represents the Hamiltonian for the interactions between the electron spin and the bath spins through the hyperfine (HF) interaction.  The electron spin Hamiltonian \(H_S\) can be expressed as
\begin{equation}
     H_{S} = \mu_BB_zg_eS_z +  D\left(S_z^2 - \frac{S(S+1)}{3}\right) + E(S_x^2 - S_y^2), \label{eq:HS}
\end{equation}
where the first term on the right-hand side of the equation describes the electron Zeeman interaction, while the second and third terms describe the zero-field splitting interaction. Here,  $\mu_B$ is Bohr magneton, $B_z$ is an external magnetic field in the direction perpendicular to the hBN, $g_e$ is a $g$ factor of an electron, $S=1$ for an electronic spin-1, $D$ is the zero-field splitting  equal to 3.479~GHz\cite{10.1038/s41524-020-0305-x} at room temperature, and $E$ is the transverse component of the ZFS, which is set to zero in this work. We note that the unit of frequency (Hz), and not the angular frequency (rad-Hz), is used throughout this paper.

\begin{figure}[ht!]
    \centering
    \includegraphics[width=0.6\linewidth]{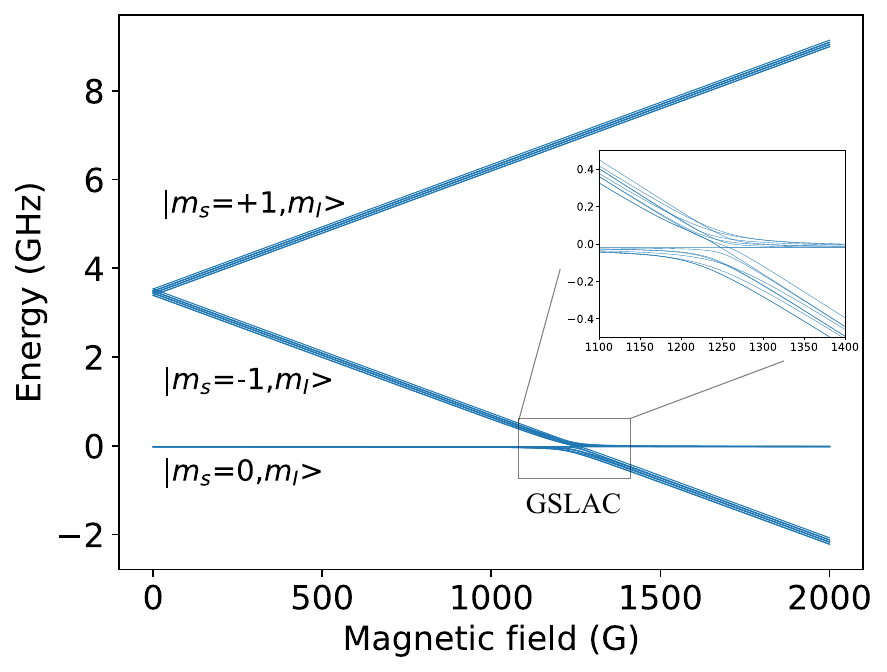}
    \caption{Magnetic field dependence of the fine structure of the ground state spin system of the V$_\text{B}^{-}$ center in hBN. Spacing of the energy levels is dominated by the zero-field splitting and Zeeman terms, except at the ground state level avoided crossing (GSLAC) of the electron-spin sublevels, where hyperfine interaction heavily mixes electron and nuclear spin states.}
    \label{fig:GSLAC}
\end{figure}

The electron spin Zeeman and zero-field splitting terms of $H_S$ are the leading terms in Eq.~\ref{eq:total_H}, which determine the main characteristics of the spectrum of the spin systems and have a major effect on the relaxation processes.
At zero magnetic field, the eigenstates corresponding to $m_s = 0$ and $m_s = \pm 1$ are split in the ground state due to the zero-field splitting. When an external magnetic field is applied, these eigenstates undergo additional splitting through the Zeeman interaction, resulting in the separation of the $\ket{m_s=\pm1, m_I}$ states, as shown in Fig.~\ref{fig:GSLAC}.

The Hamiltonian of the bath $H_{B}$ contains the following interactions
\begin{eqnarray}
    H_{B} &=& -\vec{B}\cdot\sum_{i}^{N}\gamma_i\vec{I_i}  
    + \frac{\mu_B}{4\pi}\sum_{i>j}^{N}\gamma_{i}^{n}\gamma_{j}^{n}\frac{\vec{I}_{i}\cdot\vec{I}_{j}-3(\vec{I}_{i}\cdot\hat{r}_{ij})(\vec{I}_{i}\cdot\hat{r}_{ij})}{r_{ij}^3},
    \label{eq:H_B}
\end{eqnarray}
where $\vec{I}$ is nuclear spin operator vector, $\gamma^n$ is a gyromagnetic ratio of the nuclear spins, and $N$ is the number of nuclear spins considered. Here, the first term accounts for nuclear Zeeman interaction, whereas the second term accounts for the dipole-dipole nuclear interaction, which is responsible for the nuclear-nuclear spin flip-flop process.

Finally, the interaction between the central spin and the spin bath  \(H_{\text{SB}}\) can be expressed as
\begin{eqnarray}
H_{SB} &=& \vec{S}\cdot\sum_{i}^{N}\vec{A}_{i}\cdot\vec{I}_{i}, \nonumber \\
    &=& \sum_{j=1}^{N} \bigg( A_{z z}^j S_z I_z^j+A_{x x}^j S_x I_x^j+A_{y y}^j S_y I_y^j+A_{x y}^j \left( S_x I_y^j  +  S_y I_x^j \right) \nonumber\\
     &+& A_{x z}^j \left( S_x I_z^j  +  S_z I_x^j \right) +A_{y z}^j \left( S_y I_z^j  +  S_z I_y^j \right)  \bigg), 
\label{eq:H_SB}
\end{eqnarray}
where $\vec{A}$ is the hyperfine-interaction tensor, calculated based on the method in Ref.~\cite{10.1038/s42005-024-01668-9} to tackle the finite-size effect. It is important to note that each branch in Fig.~\ref{fig:GSLAC} is further divided into four sublevels due to hyperfine interactions with the three nuclear spins. These hyperfine splittings are approximately 66 MHz for $^{15}$N. At a magnetic field of approximately 1250~G, the eigenenergies of the $\ket{m_s=0, m_I}$ and $\ket{m_s=-1, m_I}$ states approach near-degeneracy, see inset of Fig.~\ref{fig:GSLAC}. In this regime, strong mixing occurs between the electronic and nuclear spin states, enabling efficient spin flip-flop processes. This enhanced flip-flop interaction leads to pronounced correlations between spins, consequently modifying the T$_1$ relaxation behavior, as will be discussed later.

As shown in Eq.~\eqref{eq:H_SB}, the $A_{xx}$, $A_{yy}$, $A_{xy}$, and $A_{yx}$ hyperfine components govern the electron–nuclear spin flip-flop processes. To illustrate the strength of these interactions, their values are plotted in Fig.~\ref{fig:vb-1_interaction}(b), highlighting the strongly coupled nature of the V$_\text{B}^{-}$ defect system. Fig.~\ref{fig:vb-1_interaction}(b) unravels that rotational invariant perpendicular hyperfine components of the first three nearest nitrogen nuclei are strong due to the Fermi contact contribution. This suggests that relaxation via the electron–nuclear spin flip-flop process in this channel is indispensable and plays a central role in the overall spin dynamics.

To simulate spin relaxation process, our models take into account all major parts of the spin Hamiltonian in Eq.~\eqref{eq:total_H} and truncate only the nuclear spin-nuclear spin dipolar couplings to a certain degree depending on the clustering used. The dimension of the given total Hamiltonian depends on the dimension of the Hilbert space of electron and nuclear spins, and the computational demand increases exponentially with the number of nuclear spins, limiting the capability of the exact solution. To resolve this issue, the methodology developed in Ref.~\cite{10.1103/PhysRevB.101.155203} is applied by dividing the entire system into smaller clusters of spins utilizing the central spin approximation and cluster expansion to reduce the Hilbert space considered while still maintaining all the aforementioned interactions.

In short, the Hamiltonian $H_{C_k}$ of the cluster $k$ of the $N+1$ spin system  can be written as
\begin{eqnarray}
    H_{C_k} &=& H_S + \sum_{i\neq0, i\in C_k} \left(\vec{S}\vec{A}_{i}  
    -\gamma_i \vec{B} \right) \vec{I_i}  \nonumber \\
    &+& \frac{\mu_B}{4\pi}\sum_{i,j\neq i \in C_k}\gamma_{i}^{n}\gamma_{j}^{n}\frac{\vec{I}_{i}\cdot\vec{I}_{j}-3(\vec{I}_{i}\cdot\hat{r}_{ij})(\vec{I}_{i}\cdot\hat{r}_{ij})}{r_{ij}^3}.
\end{eqnarray}
We simulate the dynamics of the system through the density matrix ($\rho_{C_k}$) following the below Master equation with an extended Lindbladian developed by Ref.~\cite{10.1103/PhysRevB.101.155203}
\begin{equation}
    \frac{d\rho_{C_k}}{dt} = -\frac{i}{\hbar}[H_{C_k},\rho_{C_k}] + \mathcal{L}_{C_k}(\{\dot{b}_{C_k,mn}\},\rho_{C_k}). \label{eq:lindblad_eq}
\end{equation}
We note that $\mathcal{L}_{C_k}$ represents an \emph{extended} Lindbladian term introduced into the Master equation to account for spin flip-flop processes induced by environmental spins not included in cluster $k$ (diagonal elements in the density matrix), see Ref.~\cite{10.1103/PhysRevB.101.155203} for more details. Additionally, it is used to ensure the synchronization of key properties of different instances of the central spin across different clusters, such as $S_z$ expectation values. This correction is essential because the unsynchronization of a central spin in all clusters can lead to an artificial thermal state due to the accumulation of its magnetization. The extended  $\mathcal{L}_{C_k}$ term can be formulated as follows
\begin{equation}
\begin{aligned}
\mathcal{L}_{c_k}\left(\left\{\dot{b}_{c_k, m n}\right\}, \rho_{c_k}\right)=  \sum_{m n} \frac{\dot{b}_{c_k, m n}}{\operatorname{Tr}\left(C_{m n}^{\dagger} C_{m n} \rho_{c_k}\right)} 
 \times\left(C_{m n} \rho_{c_k} C_{m n}^{\dagger}-\frac{1}{2}\left\{\rho_{c_k}, C_{m n}^{\dagger} C_{m n}\right\}\right). \label{eq:lindbladian}
\end{aligned}
\end{equation}
Here, $\dot{b}_{C_k,mn}$ determines spin flip-flop rates of the central spin induced by environmental spins not incorporated in cluster $k$ and $C_{mn}$ is the jump operator describing the spin flip and flop transitions of the central spin between $m$ and $n$ states. As such, Eq.~\eqref{eq:lindbladian} not only causes decoherence but also contributes to the decay of the central spin.

Even though the electron spin-nuclear interactions are not approximated, the Hilbert space is significantly reduced through clusterization. In this approach, the subsystems of the central spin-spin bath model consist of only a few spins at a time, which limits the entanglement and non-classical correlations in the system that can be captured. Consequently, the choice of clusters and the definition of subsystems directly determine the accuracy of the approximation. For strongly interacting systems such as the V$_\text{B}^{-}$ defect in hBN, there is no established guideline for constructing such clusterizations. The cluster must be defined in a way that ensures the density matrix remains computationally tractable while accurately capturing the relevant correlation effects. Therefore, we investigate four different models with varying complexity and subsystem sizes to uncover relevant correlations and memory effects within the bath.

\subsection{$T_1$ Theoretical Prediction}
As detailed in the Methods section, the $T_1$ time is determined from the population of the electron spin until it relaxes to thermal equilibrium. Fig.~\ref{fig:GSLAC} illustrates the energy-level splitting among the three relevant states and suggests that the $B$ fields in the vicinity of the ground-state level anticrossing (GSLAC) at around 1250 G yield particularly narrow splitting, as the GSLAC is the point at which two ground-state sublevels become nearly degenerate. This implies that spin relaxation near GSLAC is faster, which can minimize computational expense. We therefore adopt a fixed $B$ field at 800~G to evaluate the performance of various computational models for predicting $T_1$. This value is close to, but outside, the GSLAC regime.

\begin{figure}[ht!]
\centering
    \includegraphics[width=1\linewidth]{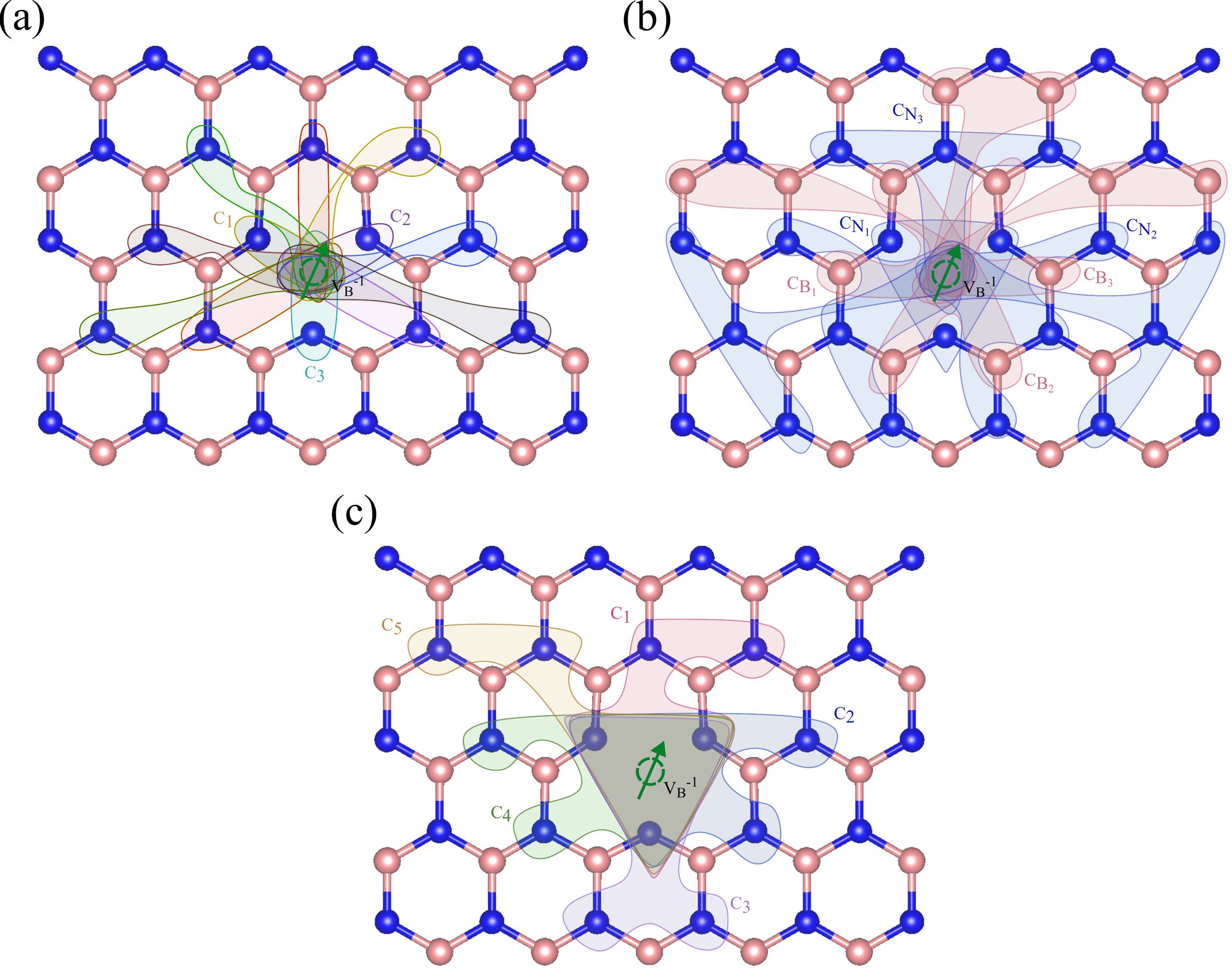}
    \caption{\textbf{Spin dynamics models.} (a) Model 1: Two-spin system model in V$_\text{B}^{-}$. Each cluster contains an electron spin and a nitrogen nuclear spin. (b) Model 2 and 3: Three/Four-spin system model in V$_\text{B}^{-}$. Blue clusters represent nitrogen clusters, containing an electron spin and three nitrogen nuclei. Pink clusters represent boron clusters, containing an electron spin and two boron nuclei. (c) Model 4: Six-spin system model in V$_\text{B}^{-}$. Each shaded area represents a six-spin cluster, including an electron spin and the closest three nitrogen nuclei in the central spin and two nitrogen nuclei from the bath. Each cluster has the same four-spin central spin and two different environmental spins. Boron nuclei are neglected. Blue and pink atoms depict nitrogen and boron atoms, respectively.}
    \label{fig:all_model}
\end{figure}

\subsection{Model 1: Two-Spin Clusters}

    \label{fig:model-1}

In this model, each cluster consists of the electron spin-1 of V$_\text{B}^{-}$ and a single nitrogen nuclear spin, as illustrated in Fig.~\ref{fig:all_model}(a). This results in a Hilbert space of dimension $\mathcal{H} = \mathcal{H}_{V_B^{-}} \otimes \mathcal{H}_{I_{^{15}N}}$, corresponding to six basis states. In total, 32 subsystems have been constructed in the cluster. The boron nuclear spins are excluded due to their weaker hyperfine interactions at this point, as shown in Fig.~\ref{fig:vb-1_interaction}(b). Based on this cluster, the dipolar nuclear interaction between nuclear spins is neglected. In addition, due to the two-spin clusterization, only electron spin-nuclear spin correlation and coherence are accounted for by the approximation.

\begin{figure*}[ht!]
    \centering
    \includegraphics[width=1\linewidth]{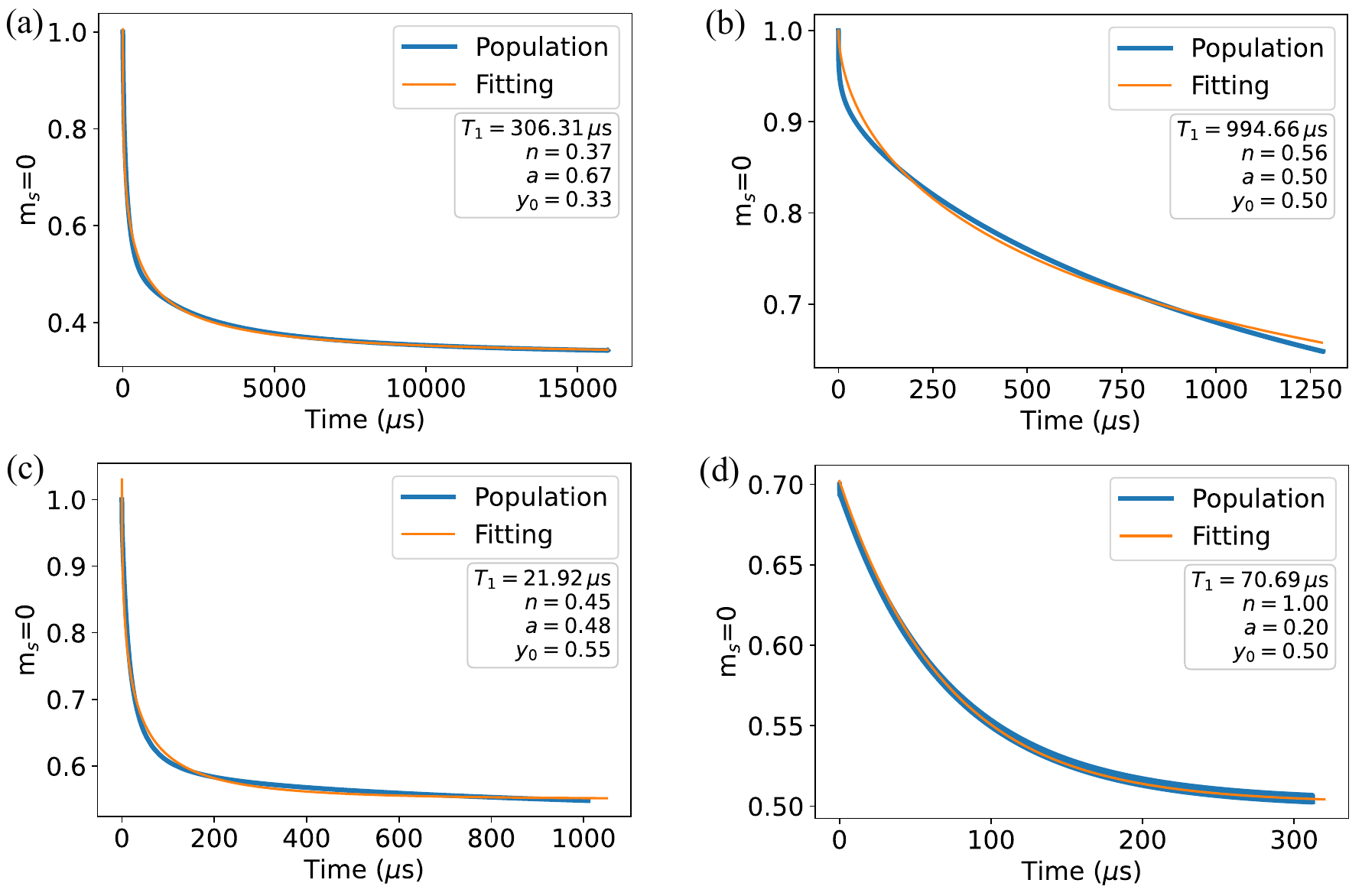}
    \caption{\textbf{Spin population of $\boldsymbol{m_s}$ = 0 state of the first cluster at $\boldsymbol{B}$ = 800 G}. The decoherence is taken into account based on (a) Model 1, (b) Model 2 with 47 nitrogen and 48 boron atom clusters, respectively, (c) Model 3 T$_2^* = 200$ ns and 47 nitrogen and 48 boron atom clusters, and (d) Model 4 with 28 nitrogen atom clusters. The $T_1$ time is obtained from fitting a$e^{(-(t/T_1)^n)}$ + $y_0$. }
    \label{fig:T1_all}
\end{figure*}

Fig.~\ref{fig:T1_all}(a) depicts the evolution of the V$_\text{B}^{-}$ electron spin initialized in the m$_s$ = 0 state. The result suggests that immediately after initialization, the electron spin polarization is rapidly transferred to the $^{15}$N nuclear spin in this model. At first, the relaxation appears exponential, but as the m$_s = 0$ population approaches the thermal equilibrium value of 1/3, as expected for the equal occupancy of the three spin sublevels (m$_s = 0, \pm1$), the relaxation pattern deviates from the exponential curve\cite{10.1103/PhysRevB.65.205309,10.1126/science.1155400} and exhibits a stretched exponential decay. This is confirmed by the exponent value of 0.37. Stretched exponents often arise when the relaxation is governed by multiple, independent decay mechanisms with different rates. While the distribution of the perpendicular hyperfine components, comprising strongly, moderately, and weakly coupled spins, see Fig.~\ref{fig:vb-1_interaction}(b), could suggest such a multi-channel relaxation mechanism, experimental results do not support this model. As a consequence, we conclude that Model 1 fails to capture relevant correlation effects of the bath that may lead to different decay mechanisms.

\subsection{Model 2 - Three/Four-Spin Clusters}

    \label{fig:model-2}
 
In this model, clusters consisting of nitrogen and boron nuclei are constructed, as depicted in Fig.~\ref{fig:all_model}(b). Each nitrogen cluster includes an electron spin-1 coupled with three nitrogen nuclear spins, whereas each boron cluster consists of an electron spin-1 coupled with only two boron nuclear spins to compensate for the larger boron spin.

To reduce the computational complexity of the density matrix, we apply a reduced density matrix approach to the electron spin-1 system. Specifically, we restrict the spin subspace to m$_s = 0$ and m$_s$ = -1, while the m$_s$ = 1 state is truncated. This approximation retains the essential spin interactions close to GSLAC while significantly (by $1/3$) reducing the density matrix size.

For nitrogen clusters, the Hilbert space dimension is  
\begin{equation}  
\mathcal{H} = \mathcal{H}_{V_B^{-},\text{reduced}} \otimes \mathcal{H}_{I_{^{15}N}} \otimes \mathcal{H}_{I_{^{15}N}} \otimes \mathcal{H}_{I_{^{15}N}} = 16,  
\end{equation}  
while for boron clusters, the Hilbert space dimension is  
\begin{equation}  
\mathcal{H} = \mathcal{H}_{V_B^{-},\text{reduced}} \otimes \mathcal{H}_{I_{^{11}B}} \otimes \mathcal{H}_{I_{^{11}B}} = 32. 
\end{equation}  

Unlike Model 1, this model incorporates dipolar interactions within each cluster, which may additionally contribute to spin dynamics behavior. In addition, Model 2 accounts for up to 4 spin correlation effects, thus, it significantly improves on Model 1 and can describe more complex dynamics. Here, 47 nitrogen clusters and 48 boron clusters have been used, describing a 238-spin model in total. Note that the number of clusters has been tested as shown in Supplementary S2.

Fig.~\ref{fig:T1_all}(b) illustrates that the electron‐spin polarization continues to be rapidly transferred to the surrounding nuclear spins until an equilibrium population. Note that the equilibrium population reaches a value of 0.5 because the reduced density matrix is restricted to the m$_s=0$ and m$_s=-1$ sublevels. Furthermore, although the relaxation decay profile appears somewhat closer to exponential than that observed in Model 1, it nonetheless deviates from the experimental exponential behavior. These observations indicate that, despite the incorporation of nuclear dipole–dipole interactions and increasing the Hilbert space of the cluster systems, the present model does not fully account for all effects necessary to reproduce the anticipated $T_1$ decay dynamics. Thus, an even more comprehensive modeling approach is required.

\subsection{Model 3 -  Three/Four-Spin Clusters with Additional Dephasing }  

Before moving toward another clusterization, we investigate the effect of decoherence of the electron spin on $T_1$ time.  The cluster Hamiltonian in this model remains the same as in Model 2, and the number of clusters is also identical. However, in this model, we incorporate the effects of spin dephasing by introducing a dephasing time (\( T_2 \)) of 200 ns \cite{10.1038/s41467-023-44494-3} into the off-diagonal terms of the Lindbladian in Eq.~\eqref{eq:lindbladian}. Note that the variation of \( T_2 \) values has been tested in Supplementary S3.

Physically, \( T_2 \) characterizes the timescale over which phase coherence between quantum states is lost due to internal noise sources. By incorporating dephasing, modeled here as an exponential decay of the magnitude of the off-diagonal elements of the reduced density matrix of the electron spin, we suppress coherent oscillations arising from Larmor precession and hyperfine-driven spin flip-flops. This, in turn, enhances the overall relaxation dynamics and shortens the \( T_1 \).

As shown in Fig.~\ref{fig:T1_all}(c), the relaxation rate (\( T^{-1} \)) increases by an order of magnitude as expected, however, the overall decay profile remains a stretched exponential and exhibits a lower decay exponent (\( n \)) compared to Model~2. These findings imply that the introduced dephasing term partially restores the behavior observed in Model~1, but it fails to accurately capture the correct relaxation dynamics.

\subsection{Model 4 - Six-Spin Clusters}
 
In this model, each cluster consists of an electron spin-1 and the three nearest-neighbor nitrogen nuclei, giving rise to an extended 4-spin central spin model, and two additional distant nitrogen nuclei, whose different in each cluster, as shown in Fig.~\ref{fig:all_model}(c). Consequently, the Hilbert space dimension is given by  
\begin{equation}  
\mathcal{H} = \mathcal{H}_{V_B^{-},\text{reduced}} \otimes \mathcal{H}_{I_{^{15}N}} \otimes \mathcal{H}_{I_{^{15}N}} \otimes \mathcal{H}_{I_{^{15}N}} \otimes \mathcal{H}_{I_{^{15}N}} \otimes \mathcal{H}_{I_{^{15}N}} = 64. 
\end{equation}  
These central and bath spin configurations enhance both the hyperfine coupling and the dipolar interaction terms in the Hamiltonian. This model accounts for all the correlation effects between the electron spin and the first neighbor nitrogen nuclear spins, as well as correlations and interactions with two additional nuclear spins. 

To manage computational complexity, we apply the same reduced density matrix approach as in the previous two models. Furthermore, a total of 28 clusters are considered in this model, accounting for a total of 60 spins. It is important to note that in the model, we focus on nitrogen nuclear spins, and boron nuclei are neglected at first, as they exhibit a weak hyperfine interaction. Furthermore, previous studies have shown that heteronuclear spin flip-flop processes, e.g., $^{29}$Si-$^{13}$C flip-flops in silicon carbide and $^{11}$B-$^{15}$N flip-flops in hBN, are strongly suppressed at nonzero $B$ fields due to the differences in the gyromagnetic ratios of the nuclear spins.\cite{10.1103/PhysRevB.90.241203} Consequently, in heterogeneous clusters consisting of the V$_\text{B}^{-}$ electron spin, the three nitrogen nuclear spins in the central spin, and additional boron atoms, no direct coupling between the two types of nuclear spins is expected. We therefore assume that the nitrogen and boron nuclear baths decouple from each other, analogous to the case of decoherence.\cite{10.1002/adfm.202511300} We note, however, that in contrast to SiC, the nuclear spins in hBN are much more strongly coupled to the electron spin, which may lead to electron-spin-mediated nuclear spin flip-flops\cite{10.1002/adfm.202511300} and potentially to additional spin relaxation mechanisms. Therefore, the computed $T_1$ time obtained from Model 4 should be regarded as an upper bound. Note that the timestep convergence test and the effect of nuclear-spin synchronization have been validated in Supplementary S4.

    \label{fig:model-4}

Model 4 enlarges the central‐spin subsystem to include the V$_\text{B}^{-}$ electron spin and its three nearest‐neighbor $^{15}$N nuclei, while the bath is augmented by two additional distant $^{15}$N nuclei. The initial electron–spin polarization in the m$_s$ = 0 sublevel is set to 70\%, in agreement with experimental measurements\cite{10.1002/qute.202300118,10.1103/PhysRevB.110.014104}. Note that the previous models initialized the electron spin polarization to 100\% in the m$_s$ = 0 state for the sake of identifying the proper model. Since Model 4 includes all relevant interactions, we initialize the electron spin according to the experimental value of 70\%. Under these conditions, the expected exponential relaxation of electron spin polarization appears, see Fig.~\ref{fig:T1_all}(d). The decay behavior is well described by an exponential function (decay exponent $n$ equal to 1). This demonstrates that the previously unaccounted correlation effects arise from insufficient entanglement and dipolar interactions within the extended central spin and also with the bath spins. That is, the spin dynamics are mediated by the spin flip-flop processes. Specifically, referring to Eqs.~\eqref{eq:H_B} and \eqref{eq:H_SB}, nuclear–nuclear flip–flop transitions are driven by the dipolar interaction term $I_i^+I_j^- + I_i^-I_j^+$ whereas electron–nuclear flip–flop transitions originate from the non‐secular hyperfine term $A_\perp(S_+I_- + S_-I_+)$. As illustrated in Fig.~\ref{fig:vb-1_interaction}(b), the V$_\text{B}^{-}$ electron spin exhibits strong hyperfine coupling to its three nearest nitrogen nuclei, therefore, including these four spins in the central‐spin description is essential to accurately capture the unseparable dynamics of the electron spin and the first neighbor nucelar spins. Simultaneously, the addition of two distant nuclear spins enables an accurate description of nuclear–nuclear flip‐flops via the dipolar interaction and additional correlation effects. Model 4 can be considered as defining the boundary between the central spin and the spin bath at the first-neighbor shell of the V$_\text{B}^{-}$ center. Consequently, the reduced bath includes only the moderately and weakly coupled nuclear spins. In this extended central spin–moderately coupled spin bath model, the spin bath can be considered ``memoryless", or at least its correlations have no significant influence on the dynamics of the extended central spin system, which, in turn, gives rise to an exponential decay characteristic of Markovian processes.
As a result, this model encompasses all significant relaxation pathways accurately and will be employed for all remaining calculations. A more detailed analysis of the contribution of distant nuclear spins is presented in the following section.

\subsection{Effect of Nuclear Spin Entanglement on $\boldsymbol{T_1}$}
\begin{figure*}[ht!]
    \centering
    \includegraphics[width=1\linewidth]{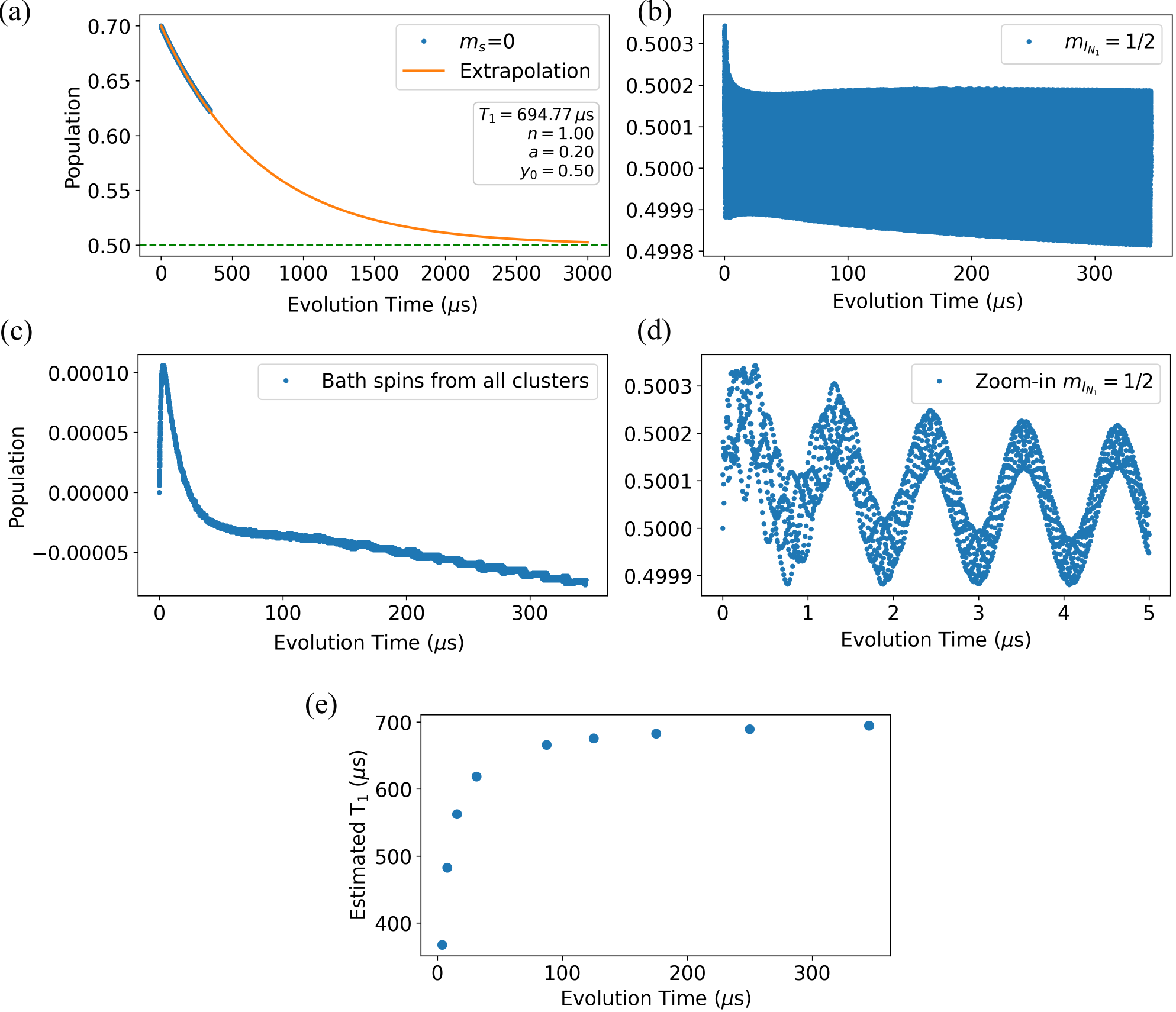}
    \caption{\textbf{Polarization dynamics at $\boldsymbol{B}$ = 90 G among 28 clusters, each containing six spins based on Model 4.} (a) Population of polarization of an electron spin of V$_\text{B}^{-}$ in m$_s$ = 0 state of the first cluster. (b)
    Population of spin-up polarization of one of the first three nearest $^{15}$N nuclear spins of the first cluster. Note that the behaviour for the other two first nearest nuclear spins remains unchanged. (c) Summation of all
    28-cluster populations of spin-up polarization of two $^{15}$N  nuclear spins
    from the bath. Note that the value of the population is shifted by 0.5 for simplicity in analyzing the spin-bath contribution. (d) The zoom-in of the nuclear-spin polarization population in (b). (e) Convergence of the extrapolated $T_1$ time as a function of evolution time.}
    \label{fig:T1_B90}
\end{figure*}

In this section, Model 4 is used to simulate the spin‐relaxation dynamics at $B = 90$ G, which is the same strength used in the experiment\cite{10.1038/s41467-023-44494-3}.
Figs.~\ref{fig:T1_B90}(a) - (c) depict the variations in a population of polarization among a single electron spin, one of the three first-nearest nuclear spins, and the sum of two distant bath spins, respectively. These results indicate that the electron spin gradually transfers its polarization to all surrounding nuclei, including both nearest neighbors and the remote bath spins, as evidenced by the sharp increase in nuclear‐spin polarization in Figs.~\ref{fig:T1_B90}(b) and (c). As time progresses, the electron‐spin polarization continues to decay until full depolarization is achieved. Both the nearest-neighbor and the more distant nuclear spins exhibit an initial transient period ($\sim$30~$\mu$s) during which the nuclear spin polarization rises to positive values before decaying. After this transient regime, the entire bath enters a steady-state period, where similar dynamics are preserved over longer timescales, leading to a gradual buildup of polarization in the bath. At a later time, not achieved in our simulations, all spin populations will converge towards a value of 0.5, indicating that the spins are depolarizing and approaching thermal equilibrium. The fast coherent oscillation of the first nearest nitrogen nuclei, as exhibited in Fig.~\ref{fig:T1_B90}(d), can be explained by the oscillation between the $\ket{m_s=0,m_I=-1/2}$ and the $\ket{m_s=-1,m_I=+1/2}$ states induced by the hyperfine interaction.

Now, turning to quantify the $T_1$ time, Fig.~\ref{fig:T1_B90}(a) showcases that the simulated polarization did not reach the thermal equilibrium point as the  simulation has reached the allowed runtime in our supercomputer, which is 20 days. We anticipated that the simulation could take up to 80 days to reach the equilibrium point. To resolve this limitation, we simulated the polarization dynamics until it reached a maximum of 20 running days. Then we fitted the curve and extrapolated the rest of the behavior. This is still valid given that the polarization will, in principle, reach the equilibrium point at the polarization equal to 0.5. Furthermore, we have cross-checked with the polarization behavior at $B$ = 800 G as shown in Fig.~\ref{fig:T1_all}(d) and found the equilibrium pattern. Note that the polarization can be simulated faster in the $B$ field close to GSLAC due to the narrower gap between $\ket{m_s=0, m_I}$ and $\ket{m_s=-1, m_I}$, as shown in Fig.~\ref{fig:GSLAC}. As a result, the fitted value of $T_1$ at $B$ = 90 G yields 694.77 $\mu$s, which is quite consistent with the experimental value of 1~ms reported at 10 K in h$^{10}$B$^{15}$N\cite{10.1038/s41467-023-44494-3}. For the small discrepancy, we hypothesize that this might arise from the absence of some correlations between different clusters, which are not captured even by our six-spin cluster approach. Moreover, while this work employed a ZFS value of 3.479 GHz, the ZFS increases to approximately 3.63 GHz at low temperatures \cite{10.1038/s41467-021-24725-1}. The increased ZFS enhances the energy separation between the three $\ket{m_s}$ states in the $0 < B < B_{\text{GSLAC}}$ interval. Consequently, the $T_1$ time is expected to be slightly larger at low magnetic fields, bringing our results into closer agreement with the experiment.

We then systematically assess the impact of the evolution time on the fitted $T_1$. As shown in Fig.~\ref{fig:T1_B90}(e), the $T_1$ converges once the simulation extends beyond 31.25 $\mu$s, at which point the fitted value of 618.7 $\mu$s underestimates the $T_1$ by 11\% (compared with the $T_1$ at a maximum evolution time of 345.2 $\mu$s, which requires 20 days of supercomputer runtime). To balance computational efficiency and accuracy in our subsequent investigation of $T_1$ as a function of magnetic field, we therefore limit all the remaining simulations to a 69 $\mu$s evolution time, which likely incurs an artifact of approximately 4.51\%.

\subsection{$\boldsymbol{T_1}$ Dependence on Magnetic Field}
\begin{figure*}[ht!]
    \centering
    \includegraphics[width=1\linewidth]{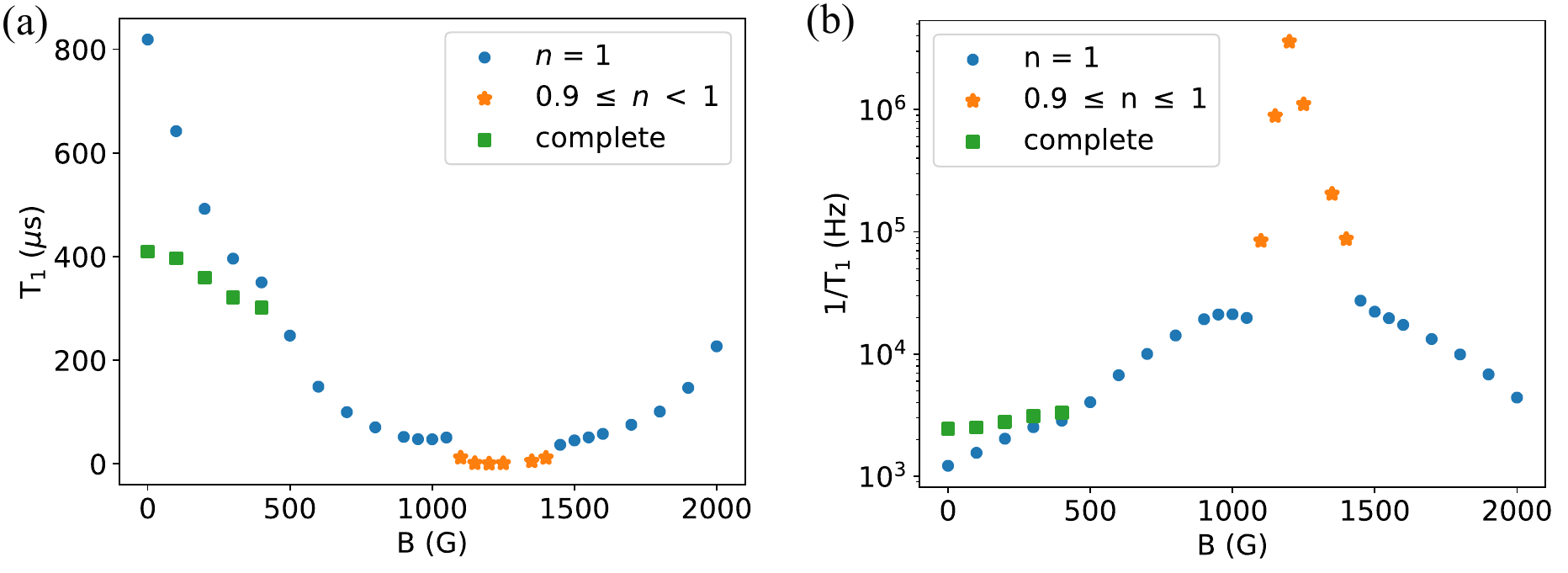}
    \caption{\textbf{The dependency of magnetic fields on decoherence.} (a) $T_1$ dependence on magnetic field. (b) The spin relaxation rate (1/$T_1$). Blue dots and orange stars indicate the $T_1$ values obtained from the single spin relaxation process via the $\ket{0}\leftrightarrow\ket{-}$ transition, using ideal exponential fitting ($n = 1$) and non-exponential fitting ($0.9 \leq n \leq 1$), respectively. Green squares represent the total $T_1$ values accounting for all relaxation pathways.}
    \label{fig:T1_varyB}
\end{figure*}

We now examine the variation of the $T_1$ time and its rate (1/$T_1$) as a function of the $B$ field, as summarized in Figs.~\ref{fig:T1_varyB}(a) and (b), respectively. Three distinct regimes can be identified: (i) the low‐field regime ($0 < B \leq 1000$ G), (ii) the near GSLAC regime ($1050 \leq B \leq 1400$ G), and (iii) the high‐field regime ($1450 \leq B \leq 2000$ G).

In low fields, $T_1$ decreases as $B$ approaches GSLAC near 1250 G. This can be attributed to the narrowing of the Zeeman splitting between $\ket{m_s = 0, m_I}$ and $\ket{m_s = -1, m_I}$ levels, as displayed in Fig.~\ref{fig:GSLAC}. The reduced energy gap enhances electron–nuclear flip–flop transitions mediated by the non‐secular hyperfine term $A_\perp(S_+I_- + S_-I_+)$, while the dipolar interaction term $I_i^+I_j^- + I_i^-I_j^+$ promotes faster nuclear–nuclear spin flip–flops by increasing bath fluctuations, as evidenced by the larger population changes of bath spins in the higher $B$ field shown in Supplementary S1. As a result, both mechanisms accelerate decoherence processes, leading to a shorter $T_1$ time with higher B strength. It is important to note that at lower fields ($B \leq 400$ G), the $T_1$ values presented in Fig.~\ref{fig:T1_varyB}(a) are likely overestimated. This is because the current model reduces the Hilbert space by applying the reduced density matrix formalism to consider only the $\ket{m_s=0}$ and $\ket{m_s=-1}$ sublevels. In practice, spin relaxation pathways involving flip-flop processes between $\ket{m_s=0}$ and $\ket{m_s=+1}$ are also possible and should be taken into account. Accordingly, the total $T_1$ should be given by the relation
\begin{equation}
    T_1 = \frac{T_1^{\ket{0}\leftrightarrow\ket{-}}T_1^{\ket{0}\leftrightarrow\ket{+1}}}{T_1^{\ket{0}\leftrightarrow\ket{-}}+T_1^{\ket{0}\leftrightarrow\ket{+1}}},
\end{equation}
where $T_1^{\ket{0}\leftrightarrow\ket{-}}$ and $T_1^{\ket{0}\leftrightarrow\ket{+1}}$ denote the relaxation times associated with spin flip-flop transitions between $\ket{m_s=0}$ and $\ket{m_s=-1}$, and between $\ket{m_s=0}$ and $\ket{m_s=+1}$, respectively. To address this limitation based on the same reduce density matrix, we computed the $T_1$ time under a $B$ field applied in the opposite direction. As shown in Fig.~\ref{fig:T1_varyB}, the total $T_1$ is slightly lower than $T_1^{\ket{0}\leftrightarrow\ket{-}}$ due to the presence of additional spin relaxation pathways. However, beyond a $B$ strength of 400 G, the total $T_1$ converges to values comparable to $T_1^{\ket{0}\leftrightarrow\ket{-}}$. This behavior arises because, at higher $B$ fields, the energy gap between $\ket{0}$ and $\ket{-}$ narrows while the gap between $\ket{0}$ and $\ket{1}$ widens, as illustrated in Fig.~\ref{fig:GSLAC}. Consequently, spin relaxation via the $\ket{0}\leftrightarrow\ket{-}$ channel dominates, and the overall relaxation rate becomes effectively determined by this single pathway. 

In the GSLAC region, the system exhibits different relaxation dynamics. Here, the \(\ket{m_s = 0, m_I}\) and \(\ket{m_s = -1, m_I}\) eigenstates undergo strong hybridization with nearby nuclear spin levels, resulting in oscillations in the population dynamics rather than the monotonic exponential decay, as shown in Supplementary S1. Under these conditions, a $T_1$ time cannot be extracted in the conventional sense based on an exponential fitting. This leads to an abrupt discontinuity in the 1/$T_1$ behavior. This behavior has also been observed in $^{13}$C in NV center in diamond\cite{10.1103/PhysRevB.101.155203}, as well as $^{13}$C and $^{29}$Si in SiC\cite{10.1038/s41524-021-00673-8}.

In high fields, although nuclear–nuclear dipolar interactions persist and continue to contribute to bath-induced noise, their influence diminishes with increasing $B$, as indicated by the reduced population fluctuations of bath spins in Supplementary S1. Furthermore, the increased energy separation between the \(\ket{m_s = 0, m_I}\) and \(\ket{m_s = -1, m_I}\) states suppresses hyperfine-mediated flip–flop transitions. Consequently, the overall relaxation rate decreases, leading to an enhancement of the $T_1$ time in this regime. It is important to note, however, that the computed $T_1$ values in this regime may still be slightly overestimated. This overestimation arises from the fact that, similar to the situation in low fields, the current model considers relaxation via the $\ket{m_s = 0} \leftrightarrow \ket{m_s = -1}$ channel while neglecting potential relaxation through the $\ket{m_s = 0} \leftrightarrow \ket{m_s = +1}$ pathway. As shown in Fig.~\ref{fig:GSLAC}, the Zeeman splitting between the $\ket{m_s = -1}$ and $\ket{m_s = +1}$ states increases substantially with $B$. At sufficiently high fields, the energy gaps between $\ket{m_s = 0}$ and both $\ket{m_s = \pm 1}$ states become approximately equal, making relaxation through both pathways equally probable. In this limit, the correct $T_1$ should be halved relative to the value obtained from a single-channel calculation; that is, $T_1 = T_1^{\ket{0}\leftrightarrow\ket{-}}/2$.

For the $B$ field beyond 2000 G, this aspect remains a computational limitation in the present work, as indicated in Supplementary S1. This is due to the very large Zeeman splitting of $\ket{m_s = -1}$ and $\ket{m_s = +1}$ at high fields; thus, capturing the oscillation over such a wide energy range requires significantly increased computational resources. Addressing this in future studies would further refine the accuracy of $T_1$ predictions in the higher field regime.

In summary, while both nuclear–nuclear dipolar interactions and electron–nuclear hyperfine coupling contribute to spin relaxation across the entire range of magnetic fields, the dominant relaxation mechanism evolves as a function of the applied field strength. At low magnetic fields, both electron–nuclear and nuclear–nuclear interactions are enhanced, leading to faster relaxation and shorter $T_1$ times. In contrast, in high fields, increasing Zeeman splitting suppresses both types of flip–flop processes, resulting in weaker interactions and prolonged $T_1$ relaxation times. However, in the vicinity of the GSLAC, the system enters a strongly coupled regime where conventional descriptions based on exponential decay fail to capture the mechanism. This therefore results in a significantly reduced $T_1$ time.

\section{Discussion}
In this work, we have developed a spin dynamics model based on the cluster expansion method to investigate the $T_1$ relaxation mechanism of the V$_\text{B}^{-}$ defect in bulk hBN. By systematically analyzing the roles of dipolar interactions and central spin–bath entanglement across four progressively refined models, we demonstrated that the V$_\text{B}^{-}$ center represents a strongly coupled spin system with its three first-neighbor nitrogen nuclear spins. This necessitates the inclusion of all three nuclei within the central spin subsystem. As a result, the central spin model comprises four spins, while the spin bath is represented by additional distant, either moderately or weakly coupled $^{15}$N nuclear spins. 

Our model was benchmarked against experimentally measured $T_1$ times, showing good agreement. The remaining discrepancies are likely due to the omission of inter-cluster correlations, which are not captured within the present 6-spin cluster expansion framework. Furthermore, we examined the dependence of $T_1$ on the external magnetic field and identified three distinct regimes: low-field, GSLAC, and high-field regions. The low- and high-field behaviors are well described by our model, while the dynamics near the GSLAC deviate from exponential relaxation due to strong electron–nuclear coupling, requiring more advanced treatments beyond the current framework.

While this work investigates the $T_1$ time in bulk hBN, dimensionality of the sample can also influence the results. In Model 4 of bulk hBN, the two distant nitrogen nuclei can sit either in the defected layer or in the neighboring layers. In contrast, in a monolayer hBN geometry, the two distant nitrogen nuclei would necessarily lie within the same plane and may therefore be located farther from the vacancy than in the bulk case. This would reduce their hyperfine interaction strength and, consequently, their contribution to decoherence. As a result, the nuclear spin bath-originated $T_1$ time in monolayer hBN is expected to be longer than in bulk or multilayer systems.

Overall, this work has established an effective and reliable model for describing $T_1$ relaxation dynamics in V$_\text{B}^{-}$ centers in hBN. The insights gained into the underlying decoherence mechanisms provide a foundation for further exploration of spin dynamics in solid-state quantum systems. For instance, future studies could extend this framework to include additional relaxation mechanisms, such as interactions with nearby paramagnetic electron spins or boron nuclei, and investigate their contributions to the overall spin relaxation behavior. However, simply replacing the two distant $^{15}$N nuclei with $^{11}$B nuclei would increase the Hilbert space from 64 to 256, making such simulations computationally demanding with the current code. Improving computational efficiency will therefore be an essential step toward including boron nuclei in a systematic way. Additionally, the work may be continued by explicitly quantifying entanglement in these strongly coupled systems. A deep understanding of spin relaxation may lead to new strategies to engineer longer $T_1$ times, either through defect environment control or external manipulation. Such advancements would be valuable for optimizing V$_\text{B}^{-}$ centers for applications in quantum sensing and quantum memory technologies.

\section{Methods} \label{sec:method}
\subsection{Computational details}
To solve the spin dynamics governed by the Lindblad master equation [Eq.~\eqref{eq:lindblad_eq}], we use our in-house MPI-parallelized C code implementing a fourth-order Runge-Kutta integration scheme. A timestep of 2.5$\times10^{-6}$ $\mu$s is employed, as validated in Supplementary S4. To simulate the dynamics of V$_\text{B}^{-}$ centers, we incorperate system-specific parameters, including the ZFS and HF interaction tensors. The ZFS is set to 3.479 GHz, as calculated using DFT together with the spin decotamination approach described in Ref.\cite{10.1038/s41524-020-0305-x}. The HF interaction is obtained from real-space calculations, which include all nuclei within a 30-\AA~ radius from the V$_\text{B}^{-}$ defect in bulk hBN to resolve the finite size effect\cite{10.1038/s42005-024-01668-9}. This corresponds to a total of 12,702 atoms in the supercell. The spin relaxation is then evaluated by monitoring the population of the V$_\text{B}^{-}$ electron spin as it evolves toward thermal equilibrium. Here, the initial density matrix is prepared such that the V$_\text{B}^{-}$ electron spin is polarized with 70\% population in the $m_s=0$ state and 30\% in the $m_s=-1$ state, while all nuclear spins are initialized in the thermal equilibrium depolarized states.

\section*{Data availability}

The data that support the findings of this study are available from the authors upon request.

\section*{Code Availability}

The codes associated with this manuscript are available from the corresponding author upon request.

\section*{Acknowledgments}
The work was supported by the National Research, Development and Innovation Office of Hungary (NKFIH) within the Quantum Information National Laboratory of Hungary (Grant No. 2022-2.1.1-NL-2022-00004) and within the project FK 145395. This project is also funded by the European Commission within Horizon Europe projects (Grant Nos.\ 101156088 and 101129663). This research is part of the Munich Quantum Valley, which is supported by the Bavarian state government with funds from the Hightech Agenda Bayern Plus. This work was funded by the Deutsche Forschungsgemeinschaft (DFG, German Research Foundation) under Germany's Excellence Strategy- EXC-2111-390814868 (MCQST) and as part of the CRC 1375 NOA project C2. The authors acknowledge support from the Federal Ministry of Research, Technology and Space (BMFTR) under grant number 13N16292 (ATOMIQS). The authors gratefully acknowledge the Gauss Centre for Supercomputing e.V.\ (www.gauss-centre.eu) for funding this project by providing computing time on the GCS Supercomputer SuperMUC-NG at Leibniz Supercomputing Centre (www.lrz.de) and on its Linux-Cluster.

\section*{Author contributions}
C.C. initiated the study and carried out the simulations. The manuscript was written by C.C. with inputs from T.V. and V.I. The work was supervised with T.V. and V.I.

\section*{Competing interests} 
The authors declare no competing interests.

\bibliography{main}

@article{xu_ab_2024,
	title = {Ab Initio Predictions of Spin Relaxation, Dephasing, and Diffusion in Solids},
	volume = {20},
	issn = {1549-9618},
	url = {https://doi.org/10.1021/acs.jctc.3c00598},
	doi = {10.1021/acs.jctc.3c00598},
	pages = {492--512},
	number = {2},
	journaltitle = {Journal of Chemical Theory and Computation},
	shortjournal = {J. Chem. Theory Comput.},
	author = {Xu, Junqing and Ping, Yuan},
	urldate = {2025-07-15},
	date = {2024-01-23},
	note = {Publisher: American Chemical Society},
}

@article{durand_optically_2023,
	title = {Optically Active Spin Defects in Few-Layer Thick Hexagonal Boron Nitride},
	volume = {131},
	url = {https://link.aps.org/doi/10.1103/PhysRevLett.131.116902},
	doi = {10.1103/PhysRevLett.131.116902},
	pages = {116902},
	number = {11},
	journaltitle = {Physical Review Letters},
	shortjournal = {Phys. Rev. Lett.},
	author = {Durand, A. and Clua-Provost, T. and Fabre, F. and Kumar, P. and Li, J. and Edgar, J. H. and Udvarhelyi, P. and Gali, A. and Marie, X. and Robert, C. and Gérard, J. M. and Gil, B. and Cassabois, G. and Jacques, V.},
	urldate = {2025-04-15},
	date = {2023-09-14},
	note = {Publisher: American Physical Society},
}

@article{ramsay_coherence_2023,
	title = {Coherence protection of spin qubits in hexagonal boron nitride},
	volume = {14},
	rights = {2023 The Author(s)},
	issn = {2041-1723},
	url = {https://www.nature.com/articles/s41467-023-36196-7},
	doi = {10.1038/s41467-023-36196-7},
	pages = {461},
	number = {1},
	journaltitle = {Nature Communications},
	shortjournal = {Nat Commun},
	author = {Ramsay, Andrew J. and Hekmati, Reza and Patrickson, Charlie J. and Baber, Simon and Arvidsson-Shukur, David R. M. and Bennett, Anthony J. and Luxmoore, Isaac J.},
	urldate = {2023-02-02},
	date = {2023-01-28},
	langid = {english},
}

@article{rizzato_extending_2023,
	title = {Extending the coherence of spin defects in {hBN} enables advanced qubit control and quantum sensing},
	volume = {14},
	rights = {2023 Springer Nature Limited},
	issn = {2041-1723},
	url = {https://www.nature.com/articles/s41467-023-40473-w},
	doi = {10.1038/s41467-023-40473-w},
	pages = {5089},
	number = {1},
	journaltitle = {Nature Communications},
	shortjournal = {Nat Commun},
	author = {Rizzato, Roberto and Schalk, Martin and Mohr, Stephan and Hermann, Jens C. and Leibold, Joachim P. and Bruckmaier, Fleming and Salvitti, Giovanna and Qian, Chenjiang and Ji, Peirui and Astakhov, Georgy V. and Kentsch, Ulrich and Helm, Manfred and Stier, Andreas V. and Finley, Jonathan J. and Bucher, Dominik B.},
	urldate = {2023-08-26},
	date = {2023-08-22},
	langid = {english},
}

@article{10.1103/PhysRevB.90.241203,
	author = {Yang, Li-Ping and Burk, Christian and Widmann, Matthias and Lee, Sang-Yun and Wrachtrup, J{\"o}rg and Zhao, Nan},
	copyright = {http://link.aps.org/licenses/aps-default-license},
	doi = {10.1103/PhysRevB.90.241203},
	issn = {1098-0121, 1550-235X},
	journal = {Phys. Rev. B},
	month = dec,
	number = {24},
	pages = {241203},
	title = {Electron spin decoherence in silicon carbide nuclear spin bath},
	url = {https://link.aps.org/doi/10.1103/PhysRevB.90.241203},
	urldate = {2025-09-12},
	volume = {90},
	year = {2014}}

@article{10.1103/PhysRevB.101.155203,
	author = {Iv{\'a}dy, Viktor},
	doi = {10.1103/PhysRevB.101.155203},
	issn = {2469-9950, 2469-9969},
	journal = {Phys. Rev. B},
	month = apr,
	number = {15},
	pages = {155203},
	title = {Longitudinal spin relaxation model applied to point-defect qubit systems},
	url = {https://link.aps.org/doi/10.1103/PhysRevB.101.155203},
	urldate = {2024-06-17},
	volume = {101},
	year = {2020}}

@article{10.1038/s41524-020-0305-x,
   author = {Viktor Ivády and Gergely Barcza and Gergő Thiering and Song Li and Hanen Hamdi and Jyh Pin Chou and Örs Legeza and Adam Gali},
   doi = {10.1038/s41524-020-0305-x},
   issn = {20573960},
   issue = {1},
   journal = {Npj Comput. Mater.},
   month ={12},
   publisher = {Nature Research},
   title = {Ab Initio Theory of the Negatively Charged Boron Vacancy Qubit in Hexagonal Boron Nitride},
   volume = {6},
   year = {2020},
   pages = {41},
url={https://www.nature.com/articles/s41524-020-0305-x}
}

@article{10.1038/s42005-024-01668-9,
	author = {Tak{\'a}cs, Istv{\'a}n and Iv{\'a}dy, Viktor},
	doi = {10.1038/s42005-024-01668-9},
	issn = {2399-3650},
	journal = {Commun Phys},
	month = jun,
	number = {1},
	pages = {178},
	shorttitle = {Accurate hyperfine tensors for solid state quantum applications},
	title = {Accurate hyperfine tensors for solid state quantum applications: case of the {NV} center in diamond},
	url = {https://www.nature.com/articles/s42005-024-01668-9},
	urldate = {2024-11-01},
	volume = {7},
	year = {2024}}

@article{10.1038/s41467-023-44494-3,
   author = {Ruotian Gong and Xinyi Du and Eli Janzen and Vincent Liu and Zhongyuan Liu and Guanghui He and Bingtian Ye and Tongcang Li and Norman Y. Yao and James H. Edgar and Erik A. Henriksen and Chong Zu},
   doi = {10.1038/s41467-023-44494-3},
   issn = {20411723},
   issue = {1},
   journal = {Nature Communications},
   month ={12},
   pmid = {38168074},
   publisher = {Nature Research},
   title = {Isotope engineering for spin defects in van der Waals materials},
   volume = {15},
   year = {2024},
   number = {104},
url = {https://www.nature.com/articles/s41467-023-44494-3}
}

@article{10.1002/qute.202300118,
	author = {Whitefield, Benjamin and Toth, Milos and Aharonovich, Igor and Tetienne, Jean‐Philippe and Kianinia, Mehran},
	doi = {10.1002/qute.202300118},
	issn = {2511-9044, 2511-9044},
	journal = {Adv Quantum Tech},
	month =jul,
	pages = {2300118},
	title = {Magnetic {Field} {Sensitivity} {Optimization} of {Negatively} {Charged} {Boron} {Vacancy} {Defects} in {hBN}.},
	year = {2023},
url = {https://advanced.onlinelibrary.wiley.com/journal/25119044}}

@article{10.1103/PhysRevB.110.014104,
  title = {Spin-dependent photodynamics of boron-vacancy centers in hexagonal boron nitride},
  author = {Clua-Provost, T. and Mu, Z. and Durand, A. and Schrader, C. and Happacher, J. and Bocquel, J. and Maletinsky, P. and Frauni\'e, J. and Marie, X. and Robert, C. and Seine, G. and Janzen, E. and Edgar, J. H. and Gil, B. and Cassabois, G. and Jacques, V.},
  journal = {Phys. Rev. B},
  volume = {110},
  issue = {1},
  pages = {014104},
  numpages = {11},
  year = {2024},
  month = {Jul},
  publisher = {American Physical Society},
  doi = {10.1103/PhysRevB.110.014104},
  url = {https://link.aps.org/doi/10.1103/PhysRevB.110.014104}
}

@article{10.1038/s41524-021-00673-8,
	author = {Bulancea-Lindvall, Oscar and Son, Nguyen T. and Abrikosov, Igor A. and Iv{\'a}dy, Viktor},
	doi = {10.1038/s41524-021-00673-8},
	issn = {2057-3960},
	journal = {npj Comput Mater},
	month = dec,
	number = {1},
	pages = {213},
	title = {Dipolar spin relaxation of divacancy qubits in silicon carbide},
	url = {https://www.nature.com/articles/s41524-021-00673-8},
	urldate = {2024-06-17},
	volume = {7},
	year = {2021}}

@article{10.1038/s41563-024-01887-z,
	author = {Stern, Hannah L. and M. Gilardoni, Carmem and Gu, Qiushi and Eizagirre Barker, Simone and Powell, Oliver F. J. and Deng, Xiaoxi and Fraser, Stephanie A. and Follet, Louis and Li, Chi and Ramsay, Andrew J. and Tan, Hark Hoe and Aharonovich, Igor and Atat{\"u}re, Mete},
	doi = {10.1038/s41563-024-01887-z},
	issn = {1476-1122, 1476-4660},
	journal = {Nat. Mater.},
	month = oct,
	number = {10},
	pages = {1379--1385},
	title = {A quantum coherent spin in hexagonal boron nitride at ambient conditions},
	url = {https://www.nature.com/articles/s41563-024-01887-z},
	urldate = {2025-04-10},
	volume = {23},
	year = {2024}}

@article{10.1038/s41467-021-24725-1,
	author = {Gottscholl, Andreas and Diez, Matthias and Soltamov, Victor and Kasper, Christian and Krau{\ss}e, Dominik and Sperlich, Andreas and Kianinia, Mehran and Bradac, Carlo and Aharonovich, Igor and Dyakonov, Vladimir},
	doi = {10.1038/s41467-021-24725-1},
	issn = {2041-1723},
	journal = {Nat Commun},
	month = jul,
	number = {1},
	pages = {4480},
	title = {Spin defects in {hBN} as promising temperature, pressure and magnetic field quantum sensors},
	url = {https://www.nature.com/articles/s41467-021-24725-1},
	urldate = {2024-09-24},
	volume = {12},
	year = {2021}}

@article{10.1016/j.physrep.2013.02.001,
	author = {Marcus W. Doherty and Neil B. Manson and Paul Delaney and Fedor Jelezko and J{\"o}rg Wrachtrup and Lloyd C.L. Hollenberg},
	doi = {https://doi.org/10.1016/j.physrep.2013.02.001},
	issn = {0370-1573},
	journal = {Physics Reports},
	number = {1},
	pages = {1-45},
	title = {The nitrogen-vacancy colour centre in diamond},
	url = {https://www.sciencedirect.com/science/article/pii/S0370157313000562},
	volume = {528},
	year = {2013}}

@article{10.1126/sciadv.abf3630,
   author = {Andreas Gottscholl and Matthias Diez and Victor Soltamov and Christian Kasper and Andreas Sperlich and Mehran Kianinia and Carlo Bradac and Igor Aharonovich and Vladimir Dyakonov},
   doi = {10.1126/sciadv.abf3630},
   issn = {2375-2548},
   issue = {14},
   journal = {Science Advances},
   month ={4},
   title = {Room temperature coherent control of spin defects in hexagonal boron nitride},
   volume = {7},
   year = {2021},
url = {https://www.science.org/doi/10.1126/sciadv.abf3630}
}

@article{10.1021/acs.nanolett.1c02495,
   author = {Xingyu Gao and Boyang Jiang and Andres E. Llacsahuanga Allcca and Kunhong Shen and Mohammad A. Sadi and Abhishek B. Solanki and Peng Ju and Zhujing Xu and Pramey Upadhyaya and Yong P. Chen and Sunil A. Bhave and Tongcang Li},
   doi = {10.1021/acs.nanolett.1c02495},
   issn = {15306992},
   journal = {Nano Letters},
   pmid = {34473524},
   publisher = {American Chemical Society},
   title = {High-Contrast Plasmonic-Enhanced Shallow Spin Defects in Hexagonal Boron Nitride for Quantum Sensing},
   year = {2021},
url = {https://pubs.acs.org/doi/10.1021/acs.nanolett.1c02495}
}

@article{10.1038/s41563-022-01329-8,
   author = {Xingyu Gao and Sumukh Vaidya and Kejun Li and Peng Ju and Boyang Jiang and Zhujing Xu and Andres E.Llacsahuanga Allcca and Kunhong Shen and Takashi Taniguchi and Kenji Watanabe and Sunil A. Bhave and Yong P. Chen and Yuan Ping and Tongcang Li},
   doi = {10.1038/s41563-022-01329-8},
   issn = {14764660},
   journal = {Nature Materials},
   publisher = {Nature Research},
   title = {Nuclear spin polarization and control in hexagonal boron nitride},
   year = {2022},
url = {https://www.nature.com/articles/s41563-022-01329-8}
}

@article{10.1038/s41563-020-0619-6,
   author = {Andreas Gottscholl and Mehran Kianinia and Victor Soltamov and Sergei Orlinskii and Georgy Mamin and Carlo Bradac and Christian Kasper and Klaus Krambrock and Andreas Sperlich and Milos Toth and Igor Aharonovich and Vladimir Dyakonov},
   doi = {10.1038/s41563-020-0619-6},
   issn = {14764660},
   issue = {5},
   journal = {Nature Materials},
   month ={5},
   pages = {540-545},
   pmid = {32094496},
   publisher = {Nature Research},
   title = {Initialization and read-out of intrinsic spin defects in a van der Waals crystal at room temperature},
   volume = {19},
   year = {2020},
url = {https://www.nature.com/articles/s41563-020-0619-6}
}

@article{10.1088/1361-6463/aa7839,
  author={Tobias Vogl and Yuerui Lu and Ping Koy Lam},
  title={Room temperature single photon source using fiber-integrated hexagonal boron nitride},
  journal={J. Phys. D: Appl. Phys.},
  volume={50},
  pages={295101},
  year={2017},
url = {https://iopscience.iop.org/article/10.1088/1361-6463/aa7839}
}

@article{10.1002/adom.202002218,
author = {Häußler, Stefan and Bayer, Gregor and Waltrich, Richard and Mendelson, Noah and Li, Chi and Hunger, David and Aharonovich, Igor and Kubanek, Alexander},
title = {Tunable Fiber-Cavity Enhanced Photon Emission from Defect Centers in {hBN}},
journal = {Adv. Opt. Mater.},
volume = {9},
pages = {2002218},
year = {2021},
url={https://advanced.onlinelibrary.wiley.com/doi/full/10.1002/adom.202002218}
}

@article{10.1038/s41467-022-33399-2,
   author = {Wei Liu and Viktor Ivády and Zhi-Peng Li and Yuan-Ze Yang and Shang Yu and Yu Meng and Zhao-An Wang and Nai-Jie Guo and Fei-Fei Yan and Qiang Li and Jun-Feng Wang and Jin-Shi Xu and Xiao Liu and Zong-Quan Zhou and Yang Dong and Xiang-Dong Chen and Fang-Wen Sun and Yi-Tao Wang and Jian-Shun Tang and Adam Gali and Chuan-Feng Li and Guang-Can Guo},
   doi = {10.1038/s41467-022-33399-2},
   issn = {2041-1723},
   issue = {1},
   journal = {Nature Communications},
   month ={9},
   pages = {5713},
   pmid = {36175507},
   publisher = {Nature Research},
   title = {Coherent dynamics of multi-spin \text{V$_\text{B}^{-1}$} center in hexagonal boron nitride},
   volume = {13},
   year = {2022},
url = {https://www.nature.com/articles/s41467-022-33399-2}
}

@article{10.1038/s41467-022-31743-0,
   author = {A. Haykal and R. Tanos and N. Minotto and A. Durand and F. Fabre and J. Li and J. H. Edgar and V. Ivády and A. Gali and T. Michel and A. Dréau and B. Gil and G. Cassabois and V. Jacques},
   doi = {10.1038/s41467-022-31743-0},
   issn = {2041-1723},
   issue = {1},
   journal = {Nature Communications},
   month ={7},
   pages = {4347},
   pmid = {35896526},
   publisher = {Nature Research},
   title = {Decoherence of \text{V$_\text{B}^{-1}$} spin defects in monoisotopic hexagonal boron nitride},
   volume = {13},
   year = {2022},
url = {https://www.nature.com/articles/s41467-022-31743-0},
}

@article{10.1038/s41524-023-01082-9,
	author = {Mondal, Sourav and Lunghi, Alessandro},
	doi = {10.1038/s41524-023-01082-9},
	issn = {2057-3960},
	journal = {npj Comput Mater},
	month = jul,
	number = {1},
	pages = {120},
	title = {Spin-phonon decoherence in solid-state paramagnetic defects from first principles},
	url = {https://www.nature.com/articles/s41524-023-01082-9},
	urldate = {2024-11-11},
	volume = {9},
	year = {2023}}

@article{10.1038/s41534-022-00637-w,
	author = {Bradley, C. E. and De Bone, S. W. and M{\"o}ller, P. F. W. and Baier, S. and Degen, M. J. and Loenen, S. J. H. and Bartling, H. P. and Markham, M. and Twitchen, D. J. and Hanson, R. and Elkouss, D. and Taminiau, T. H.},
	doi = {10.1038/s41534-022-00637-w},
	issn = {2056-6387},
	journal = {npj Quantum Inf},
	month = oct,
	number = {1},
	pages = {122},
	title = {Robust quantum-network memory based on spin qubits in isotopically engineered diamond},
	url = {https://www.nature.com/articles/s41534-022-00637-w},
	urldate = {2024-06-17},
	volume = {8},
	year = {2022}}

@article{10.1103/PhysRevX.6.021040,
	author = {Reiserer, Andreas and Kalb, Norbert and Blok, Machiel S. and Van Bemmelen, Koen J. M. and Taminiau, Tim H. and Hanson, Ronald and Twitchen, Daniel J. and Markham, Matthew},
	copyright = {http://creativecommons.org/licenses/by/3.0/},
	doi = {10.1103/PhysRevX.6.021040},
	issn = {2160-3308},
	journal = {Phys. Rev. X},
	month = jun,
	number = {2},
	pages = {021040},
	title = {Robust {Quantum}-{Network} {Memory} {Using} {Decoherence}-{Protected} {Subspaces} of {Nuclear} {Spins}},
	url = {https://link.aps.org/doi/10.1103/PhysRevX.6.021040},
	urldate = {2024-06-17},
	volume = {6},
	year = {2016}}

@article{10.1038/nphys2026,
	author = {Fuchs, G. D. and Burkard, G. and Klimov, P. V. and Awschalom, D. D.},
	copyright = {http://www.springer.com/tdm},
	doi = {10.1038/nphys2026},
	issn = {1745-2473, 1745-2481},
	journal = {Nature Phys},
	month = oct,
	number = {10},
	pages = {789--793},
	title = {A quantum memory intrinsic to single nitrogen--vacancy centres in diamond},
	url = {https://www.nature.com/articles/nphys2026},
	urldate = {2024-06-17},
	volume = {7},
	year = {2011}}

@article{10.1126/science.1220513,
	author = {Maurer, P. C. and Kucsko, G. and Latta, C. and Jiang, L. and Yao, N. Y. and Bennett, S. D. and Pastawski, F. and Hunger, D. and Chisholm, N. and Markham, M. and Twitchen, D. J. and Cirac, J. I. and Lukin, M. D.},
	doi = {10.1126/science.1220513},
	issn = {0036-8075, 1095-9203},
	journal = {Science},
	month = jun,
	number = {6086},
	pages = {1283--1286},
	title = {Room-{Temperature} {Quantum} {Bit} {Memory} {Exceeding} {One} {Second}},
	url = {https://www.science.org/doi/10.1126/science.1220513},
	urldate = {2024-06-17},
	volume = {336},
	year = {2012}}

@article{10.1063/5.0007444,
	author = {Pezzagna, S{\'e}bastien and Meijer, Jan},
	doi = {10.1063/5.0007444},
	issn = {1931-9401},
	journal = {Applied Physics Reviews},
	month = mar,
	number = {1},
	pages = {011308},
	title = {Quantum computer based on color centers in diamond},
	url = {https://pubs.aip.org/apr/article/8/1/011308/238677/Quantum-computer-based-on-color-centers-in-diamond},
	urldate = {2024-06-17},
	volume = {8},
	year = {2021}}

@article{10.1063/5.0188597,
	author = {Nateeboon, Takla and Cholsuk, Chanaprom and Vogl, Tobias and Suwanna, Sujin},
	doi = {10.1063/5.0188597},
	issn = {2835-0103},
	journal = {APL Quantum},
	month = jun,
	number = {2},
	pages = {026107},
	title = {Modeling the performance and bandwidth of single-atom adiabatic quantum memories},
	url = {https://pubs.aip.org/aip/apq/article/1/2/026107/3285615/Modeling-the-performance-and-bandwidth-of-single},
	urldate = {2024-06-17},
	volume = {1},
	year = {2024},}

@article{10.1002/adom.202302760,
	author = {Cholsuk, Chanaprom and {\c C}akan, Aslı and Suwanna, Sujin and Vogl, Tobias},
	doi = {10.1002/adom.202302760},
	issn = {2195-1071, 2195-1071},
	journal = {Advanced Optical Materials},
	month = may,
	number = {13},
	pages = {2302760},
	title = {Identifying {Electronic} {Transitions} of {Defects} in {Hexagonal} {Boron} {Nitride} for {Quantum} {Memories}},
	url = {https://onlinelibrary.wiley.com/doi/10.1002/adom.202302760},
	urldate = {2024-06-17},
	volume = {12},
	year = {2024}}

@article{10.1103/PhysRevLett.108.197601,
	author = {Jarmola, A. and Acosta, V. M. and Jensen, K. and Chemerisov, S. and Budker, D.},
	copyright = {http://link.aps.org/licenses/aps-default-license},
	doi = {10.1103/PhysRevLett.108.197601},
	issn = {0031-9007, 1079-7114},
	journal = {Phys. Rev. Lett.},
	month = may,
	number = {19},
	pages = {197601},
	title = {Temperature- and {Magnetic}-{Field}-{Dependent} {Longitudinal} {Spin} {Relaxation} in {Nitrogen}-{Vacancy} {Ensembles} in {Diamond}},
	url = {https://link.aps.org/doi/10.1103/PhysRevLett.108.197601},
	urldate = {2024-06-17},
	volume = {108},
	year = {2012}}

@article{10.1038/s41467-024-51129-8,
	author = {Scholten, Sam C. and Singh, Priya and Healey, Alexander J. and Robertson, Islay O. and Haim, Galya and Tan, Cheng and Broadway, David A. and Wang, Lan and Abe, Hiroshi and Ohshima, Takeshi and Kianinia, Mehran and Reineck, Philipp and Aharonovich, Igor and Tetienne, Jean-Philippe},
	doi = {10.1038/s41467-024-51129-8},
	issn = {2041-1723},
	journal = {Nat Commun},
	month = aug,
	number = {1},
	pages = {6727},
	title = {Multi-species optically addressable spin defects in a van der {Waals} material},
	url = {https://www.nature.com/articles/s41467-024-51129-8},
	urldate = {2025-03-25},
	volume = {15},
	year = {2024}}

@article{10.1021/acsphotonics.8b00127, title={Fabrication and Deterministic Transfer of High-Quality Quantum Emitters in Hexagonal Boron Nitride}, volume={5}, ISSN={2330-4022}, url={https://dx.doi.org/10.1021/acsphotonics.8b00127}, DOI={10.1021/acsphotonics.8b00127}, number={6}, journal={ACS Photonics}, publisher={ACS Photonics}, author={Vogl, Tobias and Campbell, Geoff and Buchler, Ben C. and Lu, Yuerui and Lam, Ping Koy}, year={2018}, pages={2305-2312}}

@article{10.1016/j.physe.2020.114251,
	author = {Wei Liu and Yi-Tao Wang and Zhi-Peng Li and Shang Yu and Zhi-Jin Ke and Yu Meng and Jian-Shun Tang and Chuan-Feng Li and Guang-Can Guo},
	doi = {https://doi.org/10.1016/j.physe.2020.114251},
	issn = {1386-9477},
	journal = {Physica E Low Dimens. Syst. Nanostruct.},
	pages = {114251},
	title = {An Ultrastable and Robust Single-Photon Emitter in Hexagonal Boron Nitride},
	url = {https://www.sciencedirect.com/science/article/pii/S1386947720306391},
	volume = {124},
	year = {2020}}

@article{10.1021/acsnano.3c08940,
	author = {Kumar, Anand and Samaner, {\c C}a{\u g}lar and Cholsuk, Chanaprom and Matthes, Tjorben and Pa{\c c}al, Serkan and Oyun, Ya{\u g}ız and Zand, Ashkan and Chapman, Robert J. and Saerens, Gr{\'e}goire and Grange, Rachel and Suwanna, Sujin and Ate{\c s}, Serkan and Vogl, Tobias},
	doi = {10.1021/acsnano.3c08940},
	journal = {ACS Nano},
	pages = {5270-5281},
	title = {Polarization Dynamics of Solid-State Quantum Emitters},
	volume = {18},
	year = {2024},
url = {https://pubs.acs.org/doi/full/10.1021/acsnano.3c08940}}

@article{10.1038/nnano.2015.242,
   author = {Toan Trong Tran and Kerem Bray and Michael J. Ford and Milos Toth and Igor Aharonovich},
   doi = {10.1038/nnano.2015.242},
   issn = {17483395},
   issue = {1},
   journal = {Nat. Nanotechnol.},
   pages = {37-41},
   pmid = {26501751},
   publisher = {Nature Publishing Group},
   title = {Quantum Emission from Hexagonal Boron Nitride Monolayers},
   volume = {11},
   year = {2016},
url = {https://www.nature.com/articles/nnano.2015.242}
}

@article{10.1126/science.1139831,
	author = {Dutt, M. V. Gurudev and Childress, L. and Jiang, L. and Togan, E. and Maze, J. and Jelezko, F. and Zibrov, A. S. and Hemmer, P. R. and Lukin, M. D.},
	doi = {10.1126/science.1139831},
	issn = {0036-8075, 1095-9203},
	journal = {Science},
	month = jun,
	number = {5829},
	pages = {1312--1316},
	title = {Quantum {Register} {Based} on {Individual} {Electronic} and {Nuclear} {Spin} {Qubits} in {Diamond}},
	url = {https://www.science.org/doi/10.1126/science.1139831},
	urldate = {2024-06-17},
	volume = {316},
	year = {2007}}

@Article{10.3390/nano12142427,
  author       = {Chanaprom Cholsuk and Sujin Suwanna and Tobias Vogl},
  date         = {2022-07},
  year = {2022},
  journal = {Nanomaterials},
  title        = {Tailoring the Emission Wavelength of Color Centers in Hexagonal Boron Nitride for Quantum Applications},
  doi          = {10.3390/nano12142427},
  number       = {14},
  pages        = {2427},
  volume       = {12},
  publisher    = {{MDPI} {AG}},
url = {https://www.mdpi.com/2079-4991/12/14/2427}
}

@article{10.1021/acs.jpclett.3c01475,
   author = {Chanaprom Cholsuk and Sujin Suwanna and Tobias Vogl},
   doi = {10.1021/acs.jpclett.3c01475},
   issn = {1948-7185},
   issue = {29},
   journal = {J. Phys. Chem. Lett.},
   month ={7},
   pages = {6564-6571},
   title = {Comprehensive Scheme for Identifying Defects in Solid-State Quantum Systems},
   volume = {14},
   year = {2023},
url = {https://pubs.acs.org/doi/10.1021/acs.jpclett.3c01475}
}

@article{10.1021/acs.jpcc.4c03404,
	author = {Cholsuk, Chanaprom and Zand, Ashkan and {\c C}akan, Aslı and Vogl, Tobias},
	copyright = {https://creativecommons.org/licenses/by/4.0/},
	doi = {10.1021/acs.jpcc.4c03404},
	issn = {1932-7447, 1932-7455},
	journal = {J. Phys. Chem. C},
	month =aug,
	number = {30},
	pages = {12716--12725},
	shorttitle = {The {hBN} {Defects} {Database}},
	title = {The {hBN} {Defects} {Database}: {A} {Theoretical} {Compilation} of {Color} {Centers} in {Hexagonal} {Boron} {Nitride}},
	volume = {128},
	year = {2024},
url = {https://pubs.acs.org/doi/10.1021/acs.jpcc.4c03404}}

@article{10.1002/adfm.202511300,
	author = {T{\'a}rk{\'a}nyi, Andr{\'a}s and Iv{\'a}dy, Viktor},
	doi = {https://doi.org/10.1002/adfm.202511300},
	eprint = {https://advanced.onlinelibrary.wiley.com/doi/pdf/10.1002/adfm.202511300},
	journal = {Advanced Functional Materials},
	pages = {e11300},
	title = {Understanding Decoherence of the Boron Vacancy Center in Hexagonal Boron Nitride},
	url = {https://advanced.onlinelibrary.wiley.com/doi/abs/10.1002/adfm.202511300}}

@article{10.1038/s41699-022-00336-2,
	author = {Lee, Jaewook and Park, Huijin and Seo, Hosung},
	doi = {10.1038/s41699-022-00336-2},
	issn = {2397-7132},
	journal = {npj 2D Mater Appl},
	month = sep,
	number = {1},
	pages = {60},
	title = {First-principles theory of extending the spin qubit coherence time in hexagonal boron nitride},
	url = {https://www.nature.com/articles/s41699-022-00336-2},
	urldate = {2024-06-17},
	volume = {6},
	year = {2022}}

@article{10.1103/PhysRevB.106.104108,
	author = {Sajid, A. and Thygesen, Kristian S.},
	doi = {10.1103/PhysRevB.106.104108},
	issn = {2469-9950, 2469-9969},
	journal = {Phys. Rev. B},
	month = sep,
	number = {10},
	pages = {104108},
	title = {Spin coherence times of point defects in two-dimensional materials from first principles},
	url = {https://link.aps.org/doi/10.1103/PhysRevB.106.104108},
	urldate = {2024-06-17},
	volume = {106},
	year = {2022}}

@article{10.1088/0034-4885/80/1/016001,
	author = {Yang, Wen and Ma, Wen-Long and Liu, Ren-Bao},
	doi = {10.1088/0034-4885/80/1/016001},
	issn = {0034-4885, 1361-6633},
	journal = {Rep. Prog. Phys.},
	month = jan,
	number = {1},
	pages = {016001},
	title = {Quantum many-body theory for electron spin decoherence in nanoscale nuclear spin baths},
	url = {https://iopscience.iop.org/article/10.1088/0034-4885/80/1/016001},
	urldate = {2024-06-17},
	volume = {80},
	year = {2017}}

@article{10.1103/PhysRevB.97.094304,
  title = {Spin-lattice relaxation of individual solid-state spins},
  author = {Norambuena, A. and Mu\~noz, E. and Dinani, H. T. and Jarmola, A. and Maletinsky, P. and Budker, D. and Maze, J. R.},
  journal = {Phys. Rev. B},
  volume = {97},
  issue = {9},
  pages = {094304},
  numpages = {13},
  year = {2018},
  month = {Mar},
  publisher = {American Physical Society},
  doi = {10.1103/PhysRevB.97.094304},
  url = {https://link.aps.org/doi/10.1103/PhysRevB.97.094304}
}

@article{10.1103/PhysRevB.98.214442,
  title = {Ab initio calculation of the spin lattice relaxation time ${T}_{1}$ for nitrogen-vacancy centers in diamond},
  author = {Gugler, J. and Astner, T. and Angerer, A. and Schmiedmayer, J. and Majer, J. and Mohn, P.},
  journal = {Phys. Rev. B},
  volume = {98},
  issue = {21},
  pages = {214442},
  numpages = {6},
  year = {2018},
  month = {Dec},
  publisher = {American Physical Society},
  doi = {10.1103/PhysRevB.98.214442},
  url = {https://link.aps.org/doi/10.1103/PhysRevB.98.214442}
}

@article{10.1103/PhysRevB.65.205309,
  title = {Electron spin relaxation by nuclei in semiconductor quantum dots},
  author = {Merkulov, I. A. and Efros, Al. L. and Rosen, M.},
  journal = {Phys. Rev. B},
  volume = {65},
  issue = {20},
  pages = {205309},
  numpages = {8},
  year = {2002},
  month = {Apr},
  publisher = {American Physical Society},
  doi = {10.1103/PhysRevB.65.205309},
  url = {https://link.aps.org/doi/10.1103/PhysRevB.65.205309}
}

@article{10.1126/science.1155400,
	author = {Hanson, R. and Dobrovitski, V. V. and Feiguin, A. E. and Gywat, O. and Awschalom, D. D.},
	doi = {10.1126/science.1155400},
	issn = {0036-8075, 1095-9203},
	journal = {Science},
	language = {en},
	month = apr,
	number = {5874},
	pages = {352--355},
	title = {Coherent {Dynamics} of a {Single} {Spin} {Interacting} with an {Adjustable} {Spin} {Bath}},
	url = {https://www.science.org/doi/10.1126/science.1155400},
	urldate = {2025-07-15},
	volume = {320},
	year = {2008}}

\end{document}